\documentclass[journal=jpcbfk,manuscript=article]{achemso}
\usepackage[version=3]{mhchem} 
\usepackage[T1]{fontenc}       

\author{M. Hvozd}
\author{T. Patsahan}
\author{M. Holovko}
\email{holovko@icmp.lviv.ua}
\affiliation[Institute for Condensed Matter Physics]
{Institute for Condensed Matter Physics of the National Academy of Sciences of Ukraine, 1 Svientsitskii St., 79011 Lviv, Ukraine}

\title[Hard Spheres and Hard Spherocylinders in a Disordered Porous Medium]
  {Isotropic-Nematic Transition and Demixing Behaviour in Binary Mixtures of Hard Spheres and Hard Spherocylinders Confined in a Disordered Porous Medium: Scaled Particle Theory}

\keywords{nematics, porous medium, phase transition, hard spheres, hard spherocylinders, binary mixture, scaled particle theory, demixing, chemical potentials}

\begin{document}


\begin{abstract}
We develop the scaled particle theory to describe the thermodynamic properties and orientation ordering 
of a binary mixture of hard spheres (HS) and hard spherocylinders (HSC) confined in a disordered porous medium. 
Using this theory the analytical expressions of the free energy, the pressure and the chemical 
potentials of HS and HSC have been derived. The improvement of obtained results is considered by introducing
the Carnahan-Starling-like and Parsons-Lee-like corrections. Phase diagrams for the isotropic-nematic transition are calculated 
from the bifurcation analysis of the integral equation for the orientation singlet distribution function 
and from the conditions of thermodynamic equilibrium. Both the approaches correctly predict the isotropic-nematic transition 
at low concentrations of hard spheres. 
However, the thermodynamic approach provides more accurate results and is able to describe the demixing phenomena in
the isotropic and nematic phases. 
The effects of porous medium on the isotropic-nematic phase transition and demixing behaviour in a binary HS/HSC mixture are discussed.
\end{abstract}

\section{Introduction}


A study of the effects of disordered porous media on the isotropic-nematic transition in a fluid of rod-like particles or elongated rigid molecules
is a topic of active research during the last decades due to the importance of such systems in numerous technological applications and because of special interest from the fundamental point of view \cite{Crawford1996}. 
For nematic fluids a porous medium can play a double role. A porous medium does not only confines a nematic fluid geometrically,
but also induces a randomly oriented field, which constrains the orientation of fluid particles near a pore surface.
The effect of such random field depends directly on the anchoring strength between fluid particles and a pore surface,
and indirectly on the porosity \cite{khasanov2005isotropic}.
Usually the starting point for the discussion of effect of this orientation field on ordering nematic phases is connected with so-called the Imry-Ma  argument \cite{imry1975random}, according to which even a low amount of static disorder leads to suppressing the nematic long-range order 
in continuous symmetry systems.
However, it was found that an existence of the quasi-long-range order can be observed in nematics confined in disordered porous media. 
Probably for the first time such a possibility was discussed in \cite{radzihovsky1997nematic}
and it was also predicted by numerical simulations \cite{chakrabarti1998simulation} as well as by a renormalization group approach \cite{feldman2000quasi}.

In many investigations of the isotropic-nematic transition of nematogenic fluids in porous media is given at the phenomenological 
or semiphenomenological levels. One of the simplest model for the description of nematic ordering in unconfined lyotropic systems is the model of hard spherocylinders (cylinders capped on both sides by hemispheres) \cite{vroege1992rep,francomelgar2008}.
The first treatment of isotropic-nematic transition within this model was done by \textit{L.~Onsager} near seventy years ago \cite{onsager1949}.
The Onsager theory is based on the low-density expansion of the free energy
functional truncated at the level of the second virial coefficient. This result is exact for the very specific case when the length of spherocylinder $L_2\rightarrow\infty$ and the diameter of spherocylinder $D_2\rightarrow 0$ are taken in such a way that the reduced density of fluid $c_2=\frac{1}{2}\pi\rho_2 L_2^2 R_2$ is fixed, where $\rho_2=N_2/V$, $N_2$ is the number of spherocylinders, $V$ is the volume of system.

The application of the scaled particle theory \cite{cotter1970statistical,cotter1974hard,cotter1978vanderwaals} provides the efficient approximate way to incorporate the higher-order density contributions neglected in the Onsager theory.
The alternative way of improvement of the Onsager theory is the Parsons-Lee (PL) approach \cite{parsons1979nematic,lee1987numerical}, which is based on the mapping of the properties of a spherocylinder fluid to those of the hard sphere model.
During the last decades the approaches developed for a hard spherocylinder fluid in the bulk case has been generalized for the description of mixtures of different hard anisotropic particles.
In such systems the new phases were observed and their properties were richer and more complicated than those for the one-component case depending on thermodynamic conditions, shapes and sizes of the components. The simplest example of such multi-component systems of hard anisotropic particles is a binary mixture of hard spheres (HS) and hard spherocylinders (HSC) for the description of which the corresponding approaches have been proposed. 
Among them there are the Onsager theory \cite{cinacchi2004liquid,vesely2005smectic},
Parsons-Lee approach \cite{malijevsky2008many,gamez2013demixing,wu2015orientational}, scaled particle theory \cite{holovko2017isotropic} and
computer simulations \cite{cuetos2007useof,malijevsky2008many,gamez2013demixing,wu2015orientational,lago2004crowding}.
The hard sphere-hard spherocylinder (HS/HSC) mixture is a simple model of a binary mixture of spherical colloids and macromolecular rod-like nematogens.
It was noticed \cite{lago2004crowding} that the properties of a HS/HSC mixture resulted from the balance between the entropic contributions of different nature. At low densities both the components are mixed in isotropic phase. At higher densities a HSC component forms the nematic phase, 
and due to a subtle balance between entropic contributions from the different components the demixing phenomena can take place, 
where the HSC and HS start to redistribute between nematic and isotropic phases \cite{cuetos2007useof,lago2004crowding}. 

In order to study fluids in a disordered matrix many different theoretical approaches \cite{given1992comment,hribar2002chemical,hribar2011partly,rosinberg2011approaches} have been developed within the model proposed by 
Madden and Glandt \cite{madden1988distribution}. According to this model the porous medium is presented as a quenched disordered matrix of hard spheres. 
Despite an intensive study of fluids confined in disordered matrices the developed approaches were numerical in their basis. 
The first attempt to obtain analytical results was done in \cite{holovko2009highly,chen2009comment}, where
the expressions for the chemical potential and pressure of a HS fluid confined in a hard sphere matrix were derived by extending 
the scaled particle theory (SPT)\cite{reiss1959statistical,reiss1959statistical,lebowitz1965scaled}.
From a subsequent improvement of the scale particle theory for a HS fluid in a hard sphere (HS) matrix the SPT2 approach was formulated in \cite{patsahan2011fluids,holovko2012fluids}.
Later on, the SPT2 approach for a HS fluid in a HS matrix was generalized for one- and two-dimensional cases \cite{holovko2010analytical}, for a fluid 
of hard convex body particles \cite{holovko2014hard}, for a multi-component mixture of hard spheres in disordered matrices \cite{chen2016scaled} and more recently for a hard spherocylinder fluid in disordered porous media \cite{holovko2015physics}.
The original SPT2 approach includes two parameters, which characterize the porosity of matrix. The first one defines a bare geometry of matrix. 
It is so-called the geometrical porosity $\phi_0$, which is equal to the ratio between the free volume not occupied by matrix particles 
and the total volume. The second parameter is so-called a probe-particle porosity, $\phi$, and is determined by the chemical potential 
of a fluid in the limit of infinite dilution. This kind of porosity characterizes the adsorption of a fluid particle in a matrix, when 
other fluid particles are absent. The porosity $\phi$ is less than $\phi_0$, since it takes into account a size of fluid particle. 
A number of approximation were proposed within the SPT2 approach\cite{patsahan2011fluids}. Among them the SPT2b approximation is considered as one of the most successful. 
It was shown that the results of SPT2b agree well with computer simulation data in a wide range of fluid densities \cite{patsahan2011fluids}.

Also it is worth mentioning that some time before we started our development of different schemes of the SPT2 theory for the description of HS fluids in disordered matrices,  \textit{M.~Schmidt} \cite{schmidt2005replica} had proposed his approach by combining the replica trick \cite{given1992comment} with the density functional theory \cite{evans1992fundamentals}, hence he had developed the replica density functional theory for the description of the thermodynamic properties of HS fluids in disordered matrices.
In particular, in \cite{schmidt2004isotropic,cheung2009quenched} the Onsager theory was generalized for the description of a HSC fluid in a hard sphere matrix. In order to control the quality of the developed theories the corresponding Monte-Carlo computer simulations were performed in those studies. 

Recently, the SPT2 theory was also extended for the description of a HSC fluid in disordered porous media \cite{holovko2014hard,holovko2015physics}. 
It was shown that the isotropic-nematic transition remains of the first order and a decrease of matrix porosity leads to the lowering of the density
of HSC particles in the coexisting phases. The obtained theoretical results are in a good agreement with computer simulation data.

Despite more or less satisfactory knowledge about the phase behaviour of HS/HSC mixturse in the bulk, our understanding of the phase behaviour of 
this binary mixture is practically absent when it is confined in a disordered porous medium. 
This paper to the best of our knowledge is the first theoretical investigation of the effect of disordered matrix on the phase behaviour of 
a HS/HSC mixture. On the basis of our recent development proposed for a hard HSC fluid in a disordered matrix \cite{holovko2014hard,holovko2015physics}, 
we generalize the SPT2 theory \cite{chen2016scaled} to describe thermodynamic properties and orientation ordering of a HS/HSC mixture. 

The paper is arranged as follows. The theoretical part is presented in Section~2. The discussion of obtained results is given in Section~3. 
And finally we draw some conclusions in the last section.

\section{Theory}
We consider a binary mixture of hard spheres (HS) and hard spherocylinders (HSC), which is confined in a disordered porous medium 
represented by the matrix of hard spheres. In order to characterize particles of the considered mixture we use three geometrical parameters:
the volume $V$ of a particle, its surface area $S$ and the mean curvature $r$ taken with a factor ${1}/{4\pi}$ \cite{holovko2015physics}.
For the HS particles with the radius $R_{1}$ these parameters are
\begin{equation}
\label{funct1}
V_{1}=\frac{4}{3}\pi R_{1}^3\;,\;\;\;   S_{1}=4\pi R_{1}^2\;,\;\;\;   r_{1}=R_{1},
\end{equation}
for the HSC particles with the radius $R_{2}$ and the length $L_{2}$
\begin{equation}
\label{funct2}
V_{2}=\pi R_{2}^2 L_{2}+\frac{4}{3}\pi R_{2}^3\;,\;\;\;   S_{2}=2\pi R_{2} L_{2}+4\pi R_{2}^2\;,\;\;\;   r_{2}=\frac{1}{4} L_{2}+R_{2}.
\end{equation}
And for the HS matrix particles with the radius $R_{0}$ we have
\begin{equation}
\label{funct0}
V_{0}=\frac{4}{3}\pi R_{0}^3\;,\;\;\;   S_{0}=4\pi R_{0}^2\;,\;\;\;   r_{0}=R_{0}.
\end{equation}

\subsection{SPT2 approach}

The main essence of the scaled particle theory (SPT) is a calculation of the work needed to insert an additional scaled particle
into a considered fluid system. A size of scaled particle is variable and in the case of HS particle 
it is defined by the scaling parameter $\lambda_{s}$. Thus, the volume $V_{1s}$, the surface area $S_{1s}$ and the mean curvature $R_{1s}$ 
of a scale particle equal to
\begin{equation}
\label{funct1s}
V_{1s}=\lambda_{s}^3 V_1,\qquad S_{1s}=\lambda_{s}^2 S_1,\qquad r_{1s}=\lambda_{s} r_1.
\end{equation}
When we insert a scaled HSC particle with the scaling radius $R_{2s}$ and the scaling length $L_{2s}$, in addition to the scaling parameter $\lambda_{s}$, we introduce the scaling parameter $\alpha_{s}$ in such a way that $R_{2s}$ and $L_{2s}$ are defined as \cite{cotter1970statistical,cotter1974hard,cotter1978vanderwaals}
\begin{equation}
\label{funct2s}
R_{2s}=\lambda_{s} R_2,\qquad L_{2s}=\alpha_{s} L_2,
\end{equation}
and consequently 
\begin{equation}
\label{funct2ss}
V_{2s}=\pi R_2^2 L_2 \alpha_s\lambda_{s}^2+\frac{4}{3} \pi R_2^3 \lambda_s^3,\quad
S_{2s}=2 \pi R_2 L_2\alpha_s\lambda_s+4\pi R_2^2 \lambda_s^2,\quad
r_{2s}=\frac{1}{4}L_2\alpha_{s}+R_2\lambda_s.
\end{equation}
Hereafter we use the conventional notations \cite{madden1988distribution,given1992comment,holovko2015physics},
where indexes ``1'' and ``2'' are used to denote HS and HSC fluid components, respectively. The index ``0'' denotes matrix particles.
For HS and HSC scaled particles we use indexes ``1s'' and ``2s'', respectively.

By inserting a scaled particle into a system we produce a cavity, which is free of fluid particles.
In the SPT theory we calculate the excess chemical potential of a scaled particle $\mu_s^{ex}$, which in turn corresponds 
to the work needed to produce the corresponding cavity \cite{reiss1959statistical,reiss1960aspects,lebowitz1965scaled}.

In the case of matrix presence we generalize the previous results for HS fluid \cite{holovko2009highly,chen2009comment,patsahan2011fluids,holovko2012fluids,chen2016scaled,holovko2017improvement}
and for fluid HSC \cite{holovko2014hard,holovko2015physics} to derive the expressions for the excess chemical potentials
of a small scaled particles in a HS/HSC mixture in the following way:
\begin{eqnarray}
\label{chem1ssmall1p}
&&\beta\mu_{1s}^{ex} (\lambda_s)=\beta\mu_{1s}-\ln(\rho_1\Lambda_1^3)=-\ln p_{01}(\lambda_s)
-\ln \bigg[1-\frac{\eta_1}{p_{01}(\lambda_s)}\left(1+\frac{r_{1s}S_1}{V_1}+\frac{r_1 S_{1s}}{V_1}+\frac{V_{1s}}{V_1}\right) \nonumber \\
&&-\frac{\eta_2}{p_{01}(\lambda_s)}\bigg(1+\frac{r_{1s}S_2}{V_2}+\frac{r_2 S_{1s}}{V_2}+\frac{V_{1s}}{V_2}\bigg)\bigg],
\end{eqnarray}
\begin{eqnarray}
\label{chem2ssmall1p}
&&\beta\mu_{2s}^{ex}(\alpha_s,\lambda_s)=\beta\mu_{2s}-\ln(\rho_2\Lambda_2^3\Lambda_{2R})=-\ln p_{02}(\alpha_s,\lambda_s) \nonumber \\
&&-\ln \bigg[1-\frac{\eta_1}{p_{02}(\alpha_s,\lambda_s)}\left(1+\frac{r_{2s}S_1}{V_1}+\frac{r_1 S_{2s}}{V_1}+\frac{V_{2s}}{V_1}\right) \nonumber \\
&&-\frac{\eta_2}{p_{02}(\alpha_s,\lambda_s)}\bigg(1+\frac{r_{2s}S_2}{V_2}+\frac{r_2 S_{2s}}{V_2}+\frac{V_{2s}}{V_2}\bigg)\bigg].
\end{eqnarray}
Here $\beta=1/k_B T$, $k_B$ is the Boltzmann constant, $T$ is the temperature, $\eta_1=\rho_1 V_1$ is the packing fraction of HS fluid, 
$\rho_1$ is the density of HS fluid, $V_1$ is the HS volume;  $\eta_2=\rho_2 V_2$ is the packing fraction of HSC fluid, 
$\rho_2$ is the density of HSC fluid, $V_2$ is the HSC volume; $\Lambda_{1}$, $\Lambda_{2}$ are the HS and HSC fluid thermal wavelengths, respectively; $\Lambda_{2R}^{-1}$ is the rotational partition function of a single HSC molecule \cite{gray1984theory}. We note that the expressions (\ref{chem1ssmall1p})-(\ref{chem2ssmall1p}) are written for isotropic case.

The terms ${p_{01}(\lambda_s)}$ and ${p_{02}(\alpha_s,\lambda_s)}$ are defined by
\begin{eqnarray}
\label{porosity01}
{p_{01}(\lambda_s)}=1-\eta_0 \left(1+\frac{r_{1s}S_0}{V_0}+\frac{r_0 S_{1s}}{V_0}+\frac{V_{1s}}{V_0}\right),
\end{eqnarray}
\begin{eqnarray}
\label{porosity02}
{p_{02}(\alpha_s,\lambda_s)}=1-\eta_0 \left(1+\frac{r_{2s}S_0}{V_0}+\frac{r_0 S_{2s}}{V_0}+\frac{V_{2s}}{V_0}\right),
\end{eqnarray}
where $\eta_0=\rho_0 V_0$ is the matrix packing fraction and $\rho_0$ is the density of matrix particles.

Substituting Eqs.~(\ref{funct1})-(\ref{funct2ss}) into Eqs.~(\ref{chem1ssmall1p})-(\ref{porosity02}) and using the generalization of Eq.~(\ref{chem2ssmall1p}) for the anisotropic case the chemical potentials of the HS and HSC scaled particles in a HS matrix can be presented as following
\begin{eqnarray}
\label{chem1ssmallp}
&&\beta\mu_{1s}^{ex} (\lambda_s)=-\ln p_{01}(\lambda_s)-\ln \bigg[1-
\frac{\eta_1}{p_{01}(\lambda_s)}(1+\lambda_s)^3 \nonumber \\
&&-\frac{\eta_2}{p_{01}(\lambda_s)} \bigg(1+\frac{1}{ k_1}\frac{6\gamma_2}{3\gamma_2-1}
\lambda_s+\frac{1}{k_1^2}\frac{3(\gamma_2+1)}{3\gamma_2-1}
\lambda_s^2+\frac{1}{k_1^3}\frac{2}{3\gamma_2-1} \lambda_s^3 \bigg)\bigg],
\end{eqnarray}
\begin{eqnarray}
\label{chem2ssmallp}
&&\beta\mu_{2s}^{ex}(\alpha_s,\lambda_s)=-\ln p_{02}(\alpha_s,\lambda_s)
-\ln \bigg[1-\frac{\eta_1}{p_{02}(\alpha_s,\lambda_s)}
\left(\frac{3}{4} s_1\alpha_s\left(1+k_1\lambda_s\right)^2+\left(1+k_1\lambda_s\right)^3\right)
\nonumber\\
&&-\frac{\eta_2}{p_{02}(\alpha_s,\lambda_s)}
\bigg(1+\frac{3(\gamma_2-1)}{3\gamma_2-1}\left[1+(\gamma_2-1)\tau(f)\right]\alpha_s
+\frac{6\gamma_2}{3\gamma_2-1}\lambda_s
\nonumber\\
&&+\frac{6(\gamma_2-1)}{3\gamma_2-1}\left[1+\frac{1}{2}(\gamma_2-1)\tau(f)\right]\alpha_s\lambda_s
+\frac{3(\gamma_2+1)}{3\gamma_2-1}\lambda_s^2 \nonumber \\
&&+\frac{3(\gamma_2-1)}{3\gamma_2-1}\alpha_s\lambda_s^2
+\frac{2}{3\gamma_2-1}\lambda_s^3\bigg)\bigg],
\end{eqnarray}
where $k_1$ and $\gamma_2$ are equal to
\begin{eqnarray}
k_1=\frac{R_2}{R_1}\;,\;\;\;\;\;\gamma_2=1+\frac{L_2}{2 R_2}.
\end{eqnarray}
In Eq.(\ref{chem2ssmallp}) $s_1$ and $\tau(f)$ are given by
\begin{equation}
s_1=\frac{L_2}{R_1},
\end{equation}
\begin{equation}
\tau(f)=\frac{4}{\pi}\int f(\Omega_1)f(\Omega_2)\sin\gamma(\Omega_1,\Omega_2)d\Omega_1d\Omega_2,
\end{equation}
where $\Omega=(\vartheta,\varphi)$ denotes the orientation of HSC particles and it is defined by the angles $\vartheta$ and $\varphi$, $d\Omega=\frac{1}{4\pi}\sin\vartheta d\vartheta d\varphi$ is the normalized angle element, $\gamma(\Omega_1, \Omega_2)$ is an angle between orientation vectors of two molecules, $f(\Omega)$ is the singlet orientation distribution function normalized in such a way that
\begin{equation}
\int f(\Omega)d\Omega=1.
\end{equation}
The term $p_{01}(\lambda_s)=\exp(-\beta\mu_{1s}^{0})$ in Eq. (\ref{chem1ssmallp}) is determined by the excess chemical potential,  
$\mu_{1s}^{0}$, of the HS scaled particle confined in an empty matrix. It has the same meaning as the probability to find a cavity 
inside of a matrix, which is large enough to insert this HS scaled particle.
Similarly, the term $p_{02}(\alpha_s,\lambda_s)=\exp(-\beta\mu_{2s}^{0})$ 
in Eq.~(\ref{chem2ssmallp}) refers to the HSC scaled particle.
For $p_{01}(\lambda_s)$ and $p_{02}(\alpha_s,\lambda_s)$ we have the following expressions:
\begin{equation}
\label{p01}
p_{01}(\lambda_s)=1-\eta_0\left(1+k_{10}\lambda_s\right)^3,
\end{equation}
where $k_{10}=R_1/R_0$ and
\begin{equation}
\label{p02}
p_{02}(\alpha_s,\lambda_s)=1-\eta_0\left(\frac{3}{4} s_0\alpha_s\left(1+k_{20}\lambda_s\right)^2+\left(1+k_{20}\lambda_s\right)^3\right),
\end{equation}
where $k_{20}=R_2/R_0$, $s_0=L_2/R_0$.

In the case of a large scaled particle we can write the excess chemical potential in the form of expression, which follows 
from the thermodynamic treatment for the work needed to produce a macroscopic cavity inside of a fluid confined in a porous medium. 
For the large HS scaled particle $\mu_{1s}^{ex} (\lambda_s)$ is related to the pressure of a HS/HSC mixture $P$ as 
\begin{equation}
\label{chem1slargep}
\beta\mu_{1s}^{ex}=w(\lambda_s)+\frac{\beta PV_{1s}}{p_{01}(\lambda_s)},
\end{equation}
where $V_{1s}$ is the volume of the HS scaled particle. For the large HSC scaled particle with the volume $V_{2s}$  we have
\begin{equation}
\label{chem2slargep}
\beta\mu_{2s}^{ex}=w(\alpha_s,\lambda_s)+\frac{\beta PV_{2s}}{p_{02}(\alpha_s,\lambda_s)}.
\end{equation}

The multipliers $1/p_{01}(\lambda_s)$ and $1/p_{02}(\alpha_s,\lambda_s)$ mean that we are dealing with excluded volumes occupied by the matrix particles. They can be considered as the probabilities of finding a cavity produced by, respectively, HS scaled particle and HSC scaled particle in the matrix when the fluid particles are absent. There are two different types of the porosities, which are related directly to these probabilities. The first type corresponds to the case of $\lambda_s=0$ and denotes the geometrical porosity
\begin{equation}
\label{geomporosity01}
p_{01}(\lambda_s=0)=\phi_{01}
\end{equation}
for the scaled HS particle and
\begin{equation}
\label{geomporosity02}
p_{02}(\alpha_s=0, \lambda_s=0)=\phi_{02}
\end{equation}
for the HSC scaled particle. The geometrical porosity is related to the volume of a void existing between matrix particles and depends only on a structure of matrix.
It is important to note that at $\alpha_s=\lambda_s=0$ the geometrical porosity $p_{01}(\lambda_s)=p_{02}(\alpha_s,\lambda_s)$, thus
\begin{equation}
\phi_{01}=\phi_{02}=\phi_0.
\end{equation}
The second type of porosity called as the probe-particle porosity is determined by the excess chemical potential of fluid particles 
in the limit of infinite dilution $\mu_\alpha^0$. Consequently, the probe particle porosity depends
on the nature of fluid under consideration. Using the SPT theory \cite{boublik1974statistical} for the bulk HS/HSC mixture in infinite dilution of corresponding component the probabilities to find HS particle or HSC particle in an empty matrix are respectively
\begin{eqnarray}
\label{probeporosity01}
&&\phi_{1}=(1-\eta_0)\exp \bigg[-3k_{10}\left(1+k_{10}\right)\frac{\eta_0}{1-\eta_0}-\frac{9}{2} k_{10}^2 \frac{\eta_0^2}{(1-\eta_0)^2} \nonumber \\
&&-k_{10}^3 \frac{\eta_0}{(1-\eta_0)^3}\left(1+\eta_0+\eta_0^2\right)\bigg],
\\
\label{probeporosity02}
&&\phi_{2}=(1-\eta_0)\exp \bigg[-3 k_{20}\left(\frac{1}{2}(\gamma_2+1)+\gamma_2 k_{20}\right) \frac{\eta_0}{1-\eta_0}
-\frac{9}{2} k_{20}^2 \gamma_2 \frac{\eta_0^2}{(1-\eta_0)^2} \nonumber \\
&&-k_{20}^3\frac{3\gamma_2-1}{2} \frac{\eta_0}{(1-\eta_0)^3} \left(1+\eta_0+\eta_0^2\right)\bigg].
\end{eqnarray}

According to the ansatz of the SPT theory
\cite{holovko2015physics,holovko2009highly,chen2009comment,patsahan2011fluids,holovko2012fluids,chen2016scaled,holovko2017improvement},
$w(\lambda_s)$ and $w(\alpha_s,\lambda_s)$ can be presented in the form of the following expansions
\begin{eqnarray}
\label{expansion1sp}
&&w(\lambda_s)=w_0+w_1\lambda_s+\frac{1}{2} w_2\lambda_s^2,
\\
\label{expansion2sp}
&&w(\alpha_s,\lambda_s)=w_{00}+w_{01}\alpha_s+w_{10}\lambda_s+w_{11}\alpha_s\lambda_s+\frac{1}{2} w_{20}\lambda_s^2.
\end{eqnarray}
The coefficients of these expansions can be obtained from the continuities of $\mu_{1s}^{ex}$ and
$\mu_{2s}^{ex}$, and they correspond to derivatives ${\partial\mu_{1s}^{ex}}/{\partial\lambda_s}$,  ${\partial^2\mu_{1s}^{ex}}/{\partial\lambda_s^2}$ at $\lambda_s=0$ for a scaled HS particle; ${\partial\mu_{2s}^{ex}}/{\partial\alpha_s}$, ${\partial\mu_{2s}^{ex}}/{\partial\lambda_s}$, ${\partial^2\mu_{2s}^{ex}}/{\partial\alpha_s\partial\lambda_s}$, ${\partial^2\mu_{2s}^{ex}}/{\partial\lambda_s^2}$ at $\alpha_s=\lambda_s=0$ for a scaled HSC particle. Thus, for a HS scaled particle we have:
\begin{eqnarray}
\label{coef1s0p}
&&w_0=-\ln\left(1-\frac{\eta}{\phi_0}\right)\nonumber,
\\
\label{coef1s1p}
&&w_1=\frac{1/\phi_0}{1-\eta/\phi_0}\bigg(3\eta_1+\frac{1}{k_1}\frac{6\gamma_2}{3\gamma_2-1}\eta_2-\frac{p'_0}{\phi_0}\eta\bigg),
\\
\label{coef1s2p}
&&w_2=\frac{1/\phi_0}{1-\eta/\phi_0}\bigg[6\eta_1+\frac{1}{k_1^2}\frac{6(\gamma_2+1)}{3\gamma_2-1}\eta_2
-2\frac{p'_0}{\phi_0}\left(3\eta_1+\frac{1}{k_1}\frac{6\gamma_2}{3\gamma_2-1}\eta_2\right)
+2\left(\frac{p'_0}{\phi_0}\right)^2\eta-\frac{p''_0}{\phi_0}\eta\bigg] \nonumber \\
&&+\left(\frac{1/\phi_0}{1-\eta/\phi_0}\right)^2
\bigg(3\eta_1+\frac{1}{k_1}\frac{6\gamma_2}{3\gamma_2-1}\eta_2-\frac{p'_0}{\phi_0}\eta\bigg)^2\nonumber.
\end{eqnarray}
Here $\eta=\eta_1+\eta_2$; $p'_0=\frac{\partial p_{01}(\lambda_s)}{\partial\lambda_s}$ and
$p''_0=\frac{\partial^2 p_{01}(\lambda_s)}{\partial\lambda_s^2}$ at $\lambda_s=0$.
For a HSC scaled particle we obtain
\begin{eqnarray}
\label{coef2s00p}
&&w_{00}=-\ln\left(1-\frac{\eta}{\phi_0}\right)\nonumber,
\\
\label{coef2s01p}
&&w_{01}=\frac{1/\phi_0}{1-\eta/\phi_0}\bigg[\frac{3}{4}s_1\eta_1
+\left(\frac{3(\gamma_2-1)}{3\gamma_2-1}+\frac{3(\gamma_2-1)^2\tau(f)}{3\gamma_2-1}\right)\eta_2
-\frac{p'_{0\alpha}}{\phi_0}\eta\bigg], \nonumber
\\
\label{coef2s10p}
&&w_{10}=\frac{1/\phi_0}{1-\eta/\phi_0}\bigg(3k_1\eta_1+\frac{6\gamma_2}{3\gamma_2-1}\eta_2
-\frac{p'_{0\lambda}}{\phi_0}\eta\bigg),  \nonumber
\\
\label{coef2s11p}
&&w_{11}=\frac{1/\phi_0}{1-\eta/\phi_0}\bigg[\frac{3}{2}k_1s_1\eta_1
+\left(\frac{6(\gamma_2-1)}{3\gamma_2-1}+\frac{3(\gamma_2-1)^2\tau(f)}{3\gamma_2-1}\right)\eta_2  \nonumber\\
&&-\frac{p'_{0\alpha}}{\phi_0}\left(3k_1\eta_1+\frac{6\gamma_2}{3\gamma_2-1}\eta_2\right)
-\frac{p'_{0\lambda}}{\phi_0}\left(\frac{3}{4}s_1\eta_1
+\left[\frac{3(\gamma_2-1)}{3\gamma_2-1}+\frac{3(\gamma_2-1)^2\tau(f)}{3\gamma_2-1}\right]\eta_2\right)
\nonumber \\
&&+2\frac{p'_{0\alpha} p'_{0\lambda}}{\phi_0^2}\eta-\frac{p''_{0\alpha\lambda}}{\phi_0}\eta\bigg]
+\left(\frac{1/\phi_0}{1-\eta/\phi_0}\right)^2
\bigg(3k_1\eta_1+\frac{6\gamma_2}{3\gamma_2-1}\eta_2-\frac{p'_{0\lambda}}{\phi_0}\eta\bigg)
\nonumber \\
&&\times
\bigg[\frac{3}{4}s_1\eta_1
+\left(\frac{3(\gamma_2-1)}{3\gamma_2-1}+\frac{3(\gamma_2-1)^2\tau(f)}{3\gamma_2-1}\right)\eta_2
-\frac{p'_{0\alpha}}{\phi_0}\eta\bigg]
\\
\label{coef2s20p}
&&w_{20}=\frac{1/\phi_0}{1-\eta/\phi_0}\bigg[6k_1^2\eta_1+\frac{6(\gamma_2+1)}{3\gamma_2-1}\eta_2
-2\frac{p'_{0\lambda}}{\phi_0}\left(3k_1\eta_1+\frac{6\gamma_2}{3\gamma_2-1}\eta_2\right)
+2\left(\frac{p'_{0\lambda}}{\phi_0}\right)^2\eta-\frac{p''_{0\lambda\lambda}}{\phi_0}\eta\bigg]
\nonumber \\
&&+\left(\frac{1/\phi_0}{1-\eta/\phi_0}\right)^2
\bigg(3k_1\eta_1+\frac{6\gamma_2}{3\gamma_2-1}\eta_2-\frac{p'_{0\lambda}}{\phi_0}\eta\bigg)^2,
\nonumber
\end{eqnarray}
where  $p'_{0\alpha}=\frac{\partial p_{02}(\alpha_s,\lambda_s)}{\partial\alpha_s}$,
$p'_{0\lambda}=\frac{\partial p_{02}(\alpha_s,\lambda_s)}{\partial\lambda_s}$,
$p''_{0\alpha\lambda}=\frac{\partial^2 p_{02}(\alpha_s,\lambda_s)}{\partial\alpha_s\partial\lambda_s}$,
$p''_{0\lambda\lambda}=\frac{\partial^2 p_{02}(\alpha_s,\lambda_s)}{\partial\lambda_s^2}$
at $\alpha_s=\lambda_s=0$.

Using Eqs.~(\ref{chem1slargep})--(\ref{chem2slargep}) at $\alpha_s=\lambda_s=1$ we derive the relations between the excess chemical potentials $\mu_1^{ex}$  and $\mu_2^{ex}$ and the pressure $P$ of a HS/HSC mixture in a matrix
\begin{eqnarray}
\label{chem1excessatlambda1p}
\beta(\mu_{1}^{ex}-\mu_1^0)=-\ln(1-\eta/\phi_0)+a_1\frac{\eta/\phi_0}{1-\eta/\phi_0}
+b_1\left(\frac{\eta/\phi_0}{1-\eta/\phi_0}\right)^2+
\frac{\beta P}{\phi_1}\frac{\eta_1}{\rho_1},
\end{eqnarray}
\begin{eqnarray}
\label{chem2excessatlambda1p}
\beta(\mu_{2}^{ex}-\mu_2^0)=-\ln(1-\eta/\phi_0)+a_2\frac{\eta/\phi_0}{1-\eta/\phi_0}
+b_2\left(\frac{\eta/\phi_0}{1-\eta/\phi_0}\right)^2+
\frac{\beta P}{\phi_2}\frac{\eta_2}{\rho_2}.
\end{eqnarray}
The coefficients $a_1$,  $a_2$, $b_1$, $b_2$ define the porous medium structure and
can be found from the following expressions:
\begin{eqnarray}
\label{a1p}
&&a_1=6\frac{\eta_1}{\eta}+\left[\frac{1}{k_1}\frac{6\gamma_2}{3\gamma_2-1}
+\frac{1}{k_1^2}\frac{3(\gamma_2+1)}{3\gamma_2-1}\right]\frac{\eta_2}{\eta}
-\frac{p'_0}{\phi_0}\left(3\frac{\eta_1}{\eta}+\frac{1}{k_1}\frac{6\gamma_2}{3\gamma_2-1}\frac{\eta_2}{\eta}\right)
\nonumber \\
&&-\frac{p'_0}{\phi_0}+\left(\frac{p'_0}{\phi_0}\right)^2-\frac{1}{2}\frac{p''_0}{\phi_0}, \nonumber
\\
\label{b1p}
&&b_1=\frac{1}{2}\bigg(3\frac{\eta_1}{\eta}+\frac{1}{k_1}\frac{6\gamma_2}{3\gamma_2-1}\frac{\eta_2}{\eta}
-\frac{p'_0}{\phi_0}\bigg)^2
\end{eqnarray}
and
\begin{eqnarray}
\label{a2p}
&&a_2(\tau(f))=\left[3k_1(1+k_1)+\frac{3}{4}s_1(1+2k_1)\right]\frac{\eta_1}{\eta}
+\left[6+\frac{6(\gamma_2-1)^2\tau(f)}{3\gamma_2-1}\right]\frac{\eta_2}{\eta} \nonumber \\
&&-\frac{p'_{0\alpha}}{\phi_0}\left(1+3k_1\frac{\eta_1}{\eta}+\frac{6\gamma_2}{3\gamma_2-1}\frac{\eta_2}{\eta}\right)
\nonumber \\
&&-\frac{p'_{0\lambda}}{\phi_0}\left[1+\left(3k_1+\frac{3}{4}s_1\right)\frac{\eta_1}{\eta}
+\left(3+\frac{3(\gamma_2-1)^2\tau(f)}{3\gamma_2-1}\right)\frac{\eta_2}{\eta}\right] \nonumber \\
&&+2\frac{p'_{0\alpha} p'_{0\lambda}}{\phi_0^2}+\left(\frac{p'_{0\lambda}}{\phi_0}\right)^2
-\frac{p''_{0\alpha\lambda}}{\phi_0}-\frac{1}{2}\frac{p''_{0\lambda\lambda}}{\phi_0}, \nonumber
\\
\label{b2p}
&&b_2(\tau(f))=\left[\left(\frac{3}{4}s_1+\frac{3}{2}k_1\right)\frac{\eta_1}{\eta}
+\left(\frac{3(2\gamma_2-1)}{3\gamma_2-1}
+\frac{3(\gamma_2-1)^2\tau(f)}{3\gamma_2-1}\right)\frac{\eta_2}{\eta}
-\frac{p'_{0\alpha}}{\phi_0}-\frac{1}{2}\frac{p'_{0\lambda}}{\phi_0}\right] \nonumber \\
&&\times \left(3k_1\frac{\eta_1}{\eta}+\frac{6\gamma_2}{3\gamma_2-1}\frac{\eta_2}{\eta}-\frac{p'_{0\lambda}}{\phi_0}\right).
\end{eqnarray}

The total chemical potentials for HS and HSC components are respectively
\begin{equation}
\label{chemtotal1p}
\beta\mu_{1}=\ln(\rho_1\Lambda_1^3)+\beta\mu_{1}^{ex}
\end{equation}
and
\begin{equation}
\label{chemtotal2p}
\beta\mu_{2}=\ln(\rho_2\Lambda_2^3\Lambda_{2R})+\beta\mu_{2}^{ex}.
\end{equation}
A substitution of Eqs.~(\ref{chem1excessatlambda1p})-(\ref{chem2excessatlambda1p})
in Eqs.~(\ref{chemtotal1p})-(\ref{chemtotal2p}) gives us two equations, and each of them contains two unknowns: 
the chemical potential and the pressure. In the case of one-component fluid we can eliminate one of unknowns, $\beta\mu_{1}$ ($\beta\mu_{2}$) or $P$, from Eq.~(\ref{chemtotal1p}) or Eq.~(\ref{chemtotal2p}) using the Gibbs-Duhem equation. In our particular case the Gibbs-Duhem equation has the
following form:
\begin{equation}
\label{GibbsDugem1p}
\frac{\partial(\beta P)}{\partial\rho}=\sum_{\alpha=1}^{2} \rho_\alpha\frac{\partial(\beta\mu_\alpha)}{\partial\rho}.
\end{equation}

Next we follow \cite{chen2016scaled} and generalize the results for a binary HS/HS mixture to the case of binary HS/HSC mixture.
In order to use Eq.(\ref{GibbsDugem1p}) and to obtain one equation containing only one unknown instead of Eqs.(\ref{chemtotal1p})-(\ref{chemtotal2p}), 
we take the derivatives with respect to the total fluid density $\rho=\sum\limits_{\alpha=1}^{2}{\rho_\alpha}$ on the both sides of Eqs.(\ref{chemtotal1p})-(\ref{chemtotal2p}) by keeping the fluid composition unchanged: $x_\alpha=\rho_\alpha/\rho$, $\alpha=1,2$. 
Hence, we can write the following
\begin{eqnarray}
\label{dchem1p}
&&\frac{\partial(\beta\mu_1)}{\partial\rho}=\frac{1}{\rho}\left[1+\frac{\eta/\phi_0}{1-\eta/\phi_0}
+a_1\frac{\eta/\phi_0}{(1-\eta/\phi_0)^2}+2b_1\frac{(\eta/\phi_0)^2}{(1-\eta/\phi_0)^3}\right] \nonumber \\
&&+\frac{4}{3}\pi R_1^3\frac{1}{\phi_1}\frac{\partial(\beta P)}{\partial\rho},
\\
\label{dchem2p}
&&\frac{\partial(\beta\mu_2)}{\partial\rho}=\frac{1}{\rho}\left[1+\frac{\eta/\phi_0}{1-\eta/\phi_0}
+a_2\frac{\eta/\phi_0}{(1-\eta/\phi_0)^2}+2b_2\frac{(\eta/\phi_0)^2}{(1-\eta/\phi_0)^3}\right] \nonumber \\
&&+\left(\pi R_2^2 L_2+\frac{4}{3}\pi R_2^3\right)
\frac{1}{\phi_2}\frac{\partial(\beta P)}{\partial\rho}.
\end{eqnarray}
Using Eq.~(\ref{GibbsDugem1p}) we can not find an expression for the chemical potential of one species from Eqs.~(\ref{dchem1p}) and (\ref{dchem2p}), 
but the combination of Eqs.~(\ref{GibbsDugem1p}) and (\ref{dchem1p})-(\ref{dchem2p}) leads to
an expression for the fluid compressibility. Taking into account that $\sum\limits_{\alpha}{x_\alpha}=1$, we obtain
\begin{eqnarray}
\label{dpressurep}
&&\frac{\partial(\beta P)}{\partial\rho}=\frac{1}{1-\eta/\phi}
+\frac{1+A}{1-\eta/\phi}\frac{\eta/\phi_0}{1-\eta/\phi_0}
+\frac{A+2B}{1-\eta/\phi}\left(\frac{\eta/\phi_0}{1-\eta/\phi_0}\right)^2 \nonumber \\
&&+\frac{2B}{1-\eta/\phi}\left(\frac{\eta/\phi_0}{1-\eta/\phi_0}\right)^3,
\end{eqnarray}
where
\begin{equation}
\label{Ap}
A=\sum_{\alpha=1}^{2}x_\alpha a_\alpha, 
\end{equation}
\begin{equation}
\label{Bp}
B=\sum_{\alpha=1}^{2}x_\alpha b_\alpha,
\end{equation}
\begin{equation}
\label{smallphip}
\frac{1}{\phi}=\frac{1}{\eta} \sum_{\alpha=1}^{2} \frac{\rho_\alpha V_\alpha}{\phi_\alpha}. 
\end{equation}
Similarly as it was done for a HS mixture in \cite{chen2016scaled} we integrate Eq.(\ref{dpressurep}) over 
the total density $\rho$ at the fixed concentration to find the pressure: 
\begin{eqnarray}
\label{pressurep}
&&\frac{\beta P}{\rho}=-\frac{\phi}{\eta} \ln\left(1-\eta/\phi\right)
+\frac{(1+A)\phi}{\phi-\phi_0}\left[\frac{\phi}{\eta} \ln\left(1-\eta/\phi\right)-
\frac{\phi_0}{\eta} \ln\left(1-\eta/\phi_0\right)\right] \nonumber \\
&&+\frac{(A+2B)\phi}{\phi-\phi_0} \bigg\{\frac{1}{1-\eta/\phi_0}
+\frac{\phi_0}{\eta} \ln\left(1-\eta/\phi_0\right) \nonumber \\
&&-\frac{\phi}{\phi-\phi_0}
\left[\frac{\phi}{\eta} \ln\left(1-\eta/\phi\right)-
\frac{\phi_0}{\eta} \ln\left(1-\eta/\phi_0\right)\right]\bigg\} \nonumber \\
&&+\frac{2B\phi}{\phi-\phi_0} \bigg\{\frac{\eta/\phi_0}{2\left(1-\eta/\phi_0\right)^2}
-\frac{1}{1-\eta/\phi_0}-\frac{\phi_0}{\eta} \ln\left(1-\eta/\phi_0\right) \nonumber \\
&&-\frac{\phi}{\phi-\phi_0}\left[\frac{1}{1-\eta/\phi_0}
+\frac{\phi_0}{\eta} \ln\left(1-\eta/\phi_0\right)\right] \nonumber \\
&&+\left(\frac{\phi}{\phi-\phi_0}\right)^2
\left[\frac{\phi}{\eta} \ln\left(1-\eta/\phi\right)-
\frac{\phi_0}{\eta} \ln\left(1-\eta/\phi_0\right)\right]\bigg\}.
\end{eqnarray}

The expression (\ref{pressurep}) is the SPT2 result for the pressure of considered mixture and it has the same form as for the one-component case \cite{patsahan2011fluids,holovko2012fluids,holovko2015physics,holovko2017improvement}.
The free energy can be obtained by integrating the pressure over the mixture density, and the chemical potentials 
we derive dy differentiating the free energy with respect to the densities of HS and HSC.

As it was noted in \cite{holovko2015physics} the obtained expression has two divergences appearing in $\eta=\phi$ and $\eta=\phi_0$. 
Since $\phi<\phi_0$ the first divergence in $\eta=\phi$ occurs at lower densities
than the second one. From geometrical point of view such a divergence should
appear at densities corresponding to the maximum value of fluid packing fraction $\eta_{max}$,
which is available for a fluid in a given matrix and should be higher than $\phi$, i.e. $\phi<\eta_{max}<\phi_0$. 
Therefore, for essentially high fluid densities the different corrections and improvements 
of SPT2 are proposed \cite{holovko2015physics,patsahan2011fluids,holovko2012fluids,chen2016scaled,holovko2017improvement}.
In the current study we restrict our consideration to the SPT2b approximation, since for mixtures in a matrix 
it is the best approximation, in which the SPT2 theory is formulated for this moment \cite{chen2016scaled}.

\subsection{SPT2b approximation for HS/HSC mixture}

We follow the scheme proposed in \cite{chen2016scaled} for a HS multi-component mixture in a matrix to derive 
the pressure and chemical potentials for a binary HS/HSC mixture using the SPT2b approximation. 
However, according to \cite{chen2016scaled} first we need to obtain the corresponding expressions 
in the SPT2a approximation \cite{patsahan2011fluids}.
As it was shown in \cite{chen2016scaled,chen2016scaled} the SPT2a approximation can be derived by replacing $\phi$ with $\phi_0$ in all 
terms of the right hand side of Eq.~(\ref{pressurep}). We take the limit $\phi\rightarrow \phi_0$ and in order to remove the singularity 
we expand $\displaystyle{\frac{\phi}{\eta}}\ln(1-\eta/\phi)$ as the Taylor series around $\phi_0$. Thus, we obtain
\begin{eqnarray}
\label{pressureSPT2a}
\left(\frac{\beta P}{\rho}\right)^{SPT2a}=\frac{1}{1-\eta/\phi_0}
+\frac{A}{2}\frac{\eta/\phi_0}{\left(1-\eta/\phi_0\right)^2}
+\frac{2B}{3}\frac{\left(\eta/\phi_0\right)^2}{\left(1-\eta/\phi_0\right)^3}.
\end{eqnarray}
From the pressure one can calculate the Helmholtz free energy using the following expression
\begin{equation}
\label{freeenergyint}
\frac{\beta F}{V}=\rho\int_{0}^{\rho} d\rho'\frac{1}{\rho'}\left(\frac{\beta P}{\rho'}\right).
\end{equation}
We carry out the integration of this expression at constant concentrations ($x_\alpha$, $\alpha=1,2$). 
Therefore, the final expression for the free energy is
\begin{eqnarray}
\label{freeenergy2a}
&&\left(\frac{\beta F}{V}\right)^{SPT2a}=\frac{\beta F_{id}}{V}-\rho_1\ln\phi_1-\rho_2\ln\phi_2 \nonumber \\
&&+\rho\bigg[-\ln(1-\eta/\phi_0)+\frac{A}{2}\frac{\eta/\phi_0}{1-\eta/\phi_0}
+\frac{B}{3}\left(\frac{\eta/\phi_0}{1-\eta/\phi_0}\right)^2\bigg],
\end{eqnarray}
where  $F_{id}$ is the Helmholtz free energy of a bulk ideal gas mixture,
\begin{equation}
\label{Fid}
\frac{\beta F_{id}}{V}=\rho_1\left[\ln(\Lambda_1^3\rho_1)-1\right]
+\rho_2\left[\ln(\Lambda_2^3\rho_2)-1\right]+\rho_2\sigma(f(\Omega)).
\end{equation}
Here $\sigma(f(\Omega))$ is the entropic term and it is defined by
\begin{equation}
\label{sigma}
\sigma(f(\Omega))=\int f(\Omega)\ln f(\Omega)d\Omega.
\end{equation}

From the Helmholtz free energy one can calculate the total chemical potentials for each of mixture components. Using the relation
\begin{equation}
\label{chemalpha}
\beta\mu_{\alpha}=\frac{\partial}{\partial\rho_{\alpha}}\left(\frac{\beta F}{V}\right),
\end{equation}
for a HS component we obtain
\begin{eqnarray}
\label{mu1Pnew}
&&(\beta\mu_1)^{SPT2a}=\ln(\Lambda_1^3\rho_1)+\beta\mu_1^0-\ln\left(1-\eta/\phi_0\right)
+\rho\frac{V_1}{\phi_0}\frac{1}{1-\eta/\phi_0} \nonumber \\
&&+\frac{1}{2}\frac{\partial}{\partial\rho_1}\left(\rho A \frac{\eta/\phi_0}{1-\eta/\phi_0}\right)
+\frac{1}{3}\frac{\partial}{\partial\rho_1}\left[\rho B \left(\frac{\eta/\phi_0}{1-\eta/\phi_0}\right)^2\right],
\end{eqnarray}
where
\begin{eqnarray}
\label{partial1A}
&&\frac{\partial}{\partial\rho_1}\left(\rho A \frac{\eta/\phi_0}{1-\eta/\phi_0}\right)=
\frac{\eta/\phi_0}{1-\eta/\phi_0}\bigg\{a_1+\frac{\rho_1 V_1}{\eta} \bigg[6-4\frac{p'_0}{\phi_0}-\frac{1}{2}\frac{p''_0}{\phi_0}+\left(\frac{p'_0}{\phi_0}\right)^2\bigg]
\nonumber \\
&&+\frac{\rho_2 V_1}{\eta}
\bigg[\frac{3}{4}s_1(1+2k_1)+3k_1(1+k_1)-\frac{p'_{0\alpha}}{\phi_0}(1+3k_1)
-\frac{p'_{0\lambda}}{\phi_0}\left(1+3k_1+\frac{3}{4}s_1\right)
\nonumber \\
&&+2\frac{p'_{0\alpha}p'_{0\lambda}}{\phi_0^2}-\frac{p''_{0\alpha\lambda}}{\phi_0}-
\frac{1}{2}\frac{p''_{0\lambda\lambda}}{\phi_0}+\left(\frac{p'_{0\lambda}}{\phi_0}\right)^2 \bigg]\bigg\}
+\rho A\frac{V_1}{\phi_0}\frac{\eta/\phi_0}{\left(1-\eta/\phi_0\right)^2}
\end{eqnarray}
and
\begin{eqnarray}
\label{partial1B}
&&\frac{\partial}{\partial\rho_1}\left[\rho B \left(\frac{\eta/\phi_0}{1-\eta/\phi_0}\right)^2\right]=
\left(\frac{\eta/\phi_0}{1-\eta/\phi_0}\right)^2
\bigg\{b_1+\frac{\rho_1 V_1}{\eta^2}\left(3-\frac{p'_0}{\phi_0}\right) \nonumber \\
&&\times \left(3\eta_1+\frac{1}{k_1}\frac{6\gamma_2}{3\gamma_2-1}\eta_2-\frac{p'_0}{\phi_0}\eta\right)
\nonumber \\
&&+\frac{\rho_2 V_1}{\eta^2}\bigg[\Big(3k_1-\frac{p'_{0\lambda}}{\phi_0}\Big)
\Big(\left[\frac{3}{2}k_1+\frac{3}{4}s_1\right]\eta_1
+\left[\frac{3(2\gamma_2-1)}{3\gamma_2-1}+\frac{3(\gamma_2-1)^2\tau(f)}{3\gamma_2-1}\right]\eta_2 \nonumber \\
&&-\frac{p'_{0\alpha}}{\phi_0}\eta-\frac{1}{2}\frac{p'_{0\lambda}}{\phi_0}\eta\Big)
\nonumber \\
&&+\left(\frac{3}{2}k_1+\frac{3}{4}s_1-\frac{p'_{0\alpha}}{\phi_0}-\frac{1}{2}\frac{p'_{0\lambda}}{\phi_0}\right)
\left(3k_1\eta_1+\frac{6\gamma_2}{3\gamma_2-1}\eta_2-\frac{p'_{0\lambda}}{\phi_0}\eta\right)\bigg]\bigg\} \nonumber \\
&&+2\rho B \frac{V_1}{\phi_0}\frac{\left(\eta/\phi_0\right)^2}{\left(1-\eta/\phi_0\right)^3}.
\end{eqnarray}
For a HSC component we have the following expression
\begin{eqnarray}
\label{mu2Pnew}
&&(\beta\mu_2)^{SPT2a}=\ln(\Lambda_2^3\rho_2)+\sigma(f(\Omega))+\beta\mu_2^0-\ln\left(1-\eta/\phi_0\right)
+\rho\frac{V_2}{\phi_0}\frac{1}{1-\eta/\phi_0} \nonumber \\
&&+\frac{1}{2}\frac{\partial}{\partial\rho_2}\left(\rho A \frac{\eta/\phi_0}{1-\eta/\phi_0}\right)
+\frac{1}{3}\frac{\partial}{\partial\rho_2}\left[\rho B \left(\frac{\eta/\phi_0}{1-\eta/\phi_0}\right)^2\right],
\end{eqnarray}
where
\begin{eqnarray}
\label{partial2A}
&&\frac{\partial}{\partial\rho_2}\left(\rho A \frac{\eta/\phi_0}{1-\eta/\phi_0}\right)=
\frac{\eta/\phi_0}{1-\eta/\phi_0}\bigg\{a_2+\frac{\rho_1 V_2}{\eta}\bigg[\frac{1}{k_1}\frac{6\gamma_2}{3\gamma_2-1}+\frac{1}{k_1^2}\frac{3(\gamma_2+1)}{3\gamma_2-1} \nonumber \\
&&-\frac{p'_0}{\phi_0}\frac{1}{k_1}\frac{6\gamma_2}{3\gamma_2-1}
-\frac{p'_0}{\phi_0}-\frac{1}{2}\frac{p''_0}{\phi_0}+\left(\frac{p'_0}{\phi_0}\right)^2\bigg] \nonumber \\
&&+\frac{\rho_2 V_2}{\eta}
\bigg[6+\frac{6(\gamma_2-1)^2\tau(f)}{3\gamma_2-1}
-\frac{p'_{0\alpha}}{\phi_0}\left(1+\frac{6\gamma_2}{3\gamma_2-1}\right) \nonumber \\
&&-\frac{p'_{0\lambda}}{\phi_0}\left(4+\frac{3(\gamma_2-1)^2\tau(f)}{3\gamma_2-1}\right)
+2\frac{p'_{0\alpha}p'_{0\lambda}}{\phi_0^2}-\frac{p''_{0\alpha\lambda}}{\phi_0}
-\frac{1}{2}\frac{p''_{0\lambda\lambda}}{\phi_0}+\left(\frac{p'_{0\lambda}}{\phi_0}\right)^2\bigg]\bigg\} \nonumber \\
&&+\rho A\frac{V_2}{\phi_0}\frac{\eta/\phi_0}{\left(1-\eta/\phi_0\right)^2}
\end{eqnarray}
and
\begin{eqnarray}
\label{partial2B}
&&\frac{\partial}{\partial\rho_2}\left[\rho B \left(\frac{\eta/\phi_0}{1-\eta/\phi_0}\right)^2\right]=
\left(\frac{\eta/\phi_0}{1-\eta/\phi_0}\right)^2
\bigg\{b_2+\frac{\rho_1 V_2}{\eta^2}\left(\frac{1}{k_1}\frac{6\gamma_2}{3\gamma_2-1}-\frac{p'_0}{\phi_0}\right) \nonumber \\
&&\times\left(3\eta_1+\frac{1}{k_1}\frac{6\gamma_2}{3\gamma_2-1}\eta_2-\frac{p'_0}{\phi_0}\eta\right)
+\frac{\rho_2 V_2}{\eta^2}
\bigg[\left(\frac{6\gamma_2}{3\gamma_2-1}-\frac{p'_{0\lambda}}{\phi_0}\right)
\bigg(\left[\frac{3}{2}k_1+\frac{3}{4}s_1\right]\eta_1 \nonumber \\
&&+\left[\frac{3(2\gamma_2-1)}{3\gamma_2-1}+\frac{3(\gamma_2-1)^2\tau(f)}{3\gamma_2-1}\right]\eta_2
-\frac{p'_{0\alpha}}{\phi_0}\eta-\frac{1}{2}\frac{p'_{0\lambda}}{\phi_0}\eta\bigg) \nonumber \\
&&+\left(\frac{3(2\gamma_2-1)}{3\gamma_2-1}+\frac{3(\gamma_2-1)^2\tau(f)}{3\gamma_2-1}
-\frac{p'_{0\alpha}}{\phi_0}-\frac{1}{2}\frac{p'_{0\lambda}}{\phi_0}\right)
\left(3k_1\eta_1+\frac{6\gamma_2}{3\gamma_2-1}\eta_2-\frac{p'_{0\lambda}}{\phi_0}\eta\right)\bigg]\bigg\} \nonumber \\
&&+2\rho B\frac{V_2}{\phi_0}\frac{\left(\eta/\phi_0\right)^2}{\left(1-\eta/\phi_0\right)^3}.
\end{eqnarray}

Now, we take the limit $\phi\rightarrow\phi_0$ in all terms of Eq.~(\ref{pressurep}) except the first term,
and as a result, the pressure in the SPT2b approximation is obtained:
\begin{eqnarray}
\label{pressureSPT2b}
\left(\frac{\beta P}{\rho}\right)^{SPT2b}=\left(\frac{\beta P}{\rho}\right)^{SPT2a}
-\frac{\phi}{\eta} \ln\left(1-\eta/\phi\right)+\frac{\phi_0}{\eta} \ln\left(1-\eta/\phi_0\right).
\end{eqnarray}
Consequently, for the free energy we have
\begin{eqnarray}
\label{F2b}
\left(\frac{\beta F}{V}\right)^{SPT2b}=\left(\frac{\beta F}{V}\right)^{SPT2a}
+\rho\bigg\{-\left(1-\frac{\phi}{\eta}\right)\ln(1-\eta/\phi)
+\left(1-\frac{\phi_0}{\eta}\right)\ln(1-\eta/\phi_0)\bigg\}.
\end{eqnarray}
The total chemical potentials for the both of components of a HS/HSC mixture are derived in the SPT2b approximation as well:
\begin{eqnarray}
\label{chemSPT2b}
&&\beta\mu_\alpha^{SPT2b}=\beta\mu_\alpha^{SPT2a}-\ln\left(1-\eta/\phi\right)+\ln\left(1-\eta/\phi_0\right)
+\left(\frac{\rho V_\alpha}{\eta}-1\right)\frac{\beta(P^{SPT2b}-P^{SPT2a})}{\rho} \nonumber \\
&&-\frac{\rho V_\alpha}{\eta}\left(\frac{\phi}{\phi_\alpha}-1\right)
\left[\frac{\phi}{\eta}\ln\left(1-\eta/\phi\right)+1\right].
\end{eqnarray}

In order to obtain the singlet orientation distribution function $f(\Omega)$ a minimization of the free energy 
with respect to variations of $f(\Omega)$ should be applied. After taking the corresponding functional derivation 
we come to the non-linear integral equation:
\begin{equation}
\label{nonlineareq}
\ln f(\Omega_1)+\lambda+\frac{8}{\pi}C\int f(\Omega')\sin\gamma(\Omega_1\Omega')d\Omega'=0,
\end{equation}
where
\begin{eqnarray}
\label{Cp}
&&C=\frac{\eta_2/\phi_0}{1-\eta/\phi_0}
\bigg[\frac{3(\gamma_2-1)^2}{3\gamma_2-1}\left(1-\frac{1}{2}\frac{p'_{0\lambda}}{\phi_0}\right) \nonumber \\
&&+\frac{1/\phi_0}{1-\eta/\phi_0}\frac{(\gamma_2-1)^2}{3\gamma_2-1}
\left(3k_1\eta_1+\frac{6\gamma_2}{3\gamma_2-1}\eta_2-\frac{p'_{0\lambda}}{\phi_0}\eta\right)\bigg],
\end{eqnarray}
\begin{equation}
\label{phi_0}
\phi_0=1-\eta_0,
\end{equation}
\begin{equation}
\label{p0lambda}
p'_{0\lambda}=-3\eta_0 k_{20}.
\end{equation}

\subsection{Carnahan-Starling and Parsons-Lee corrections}

In our previous paper \cite{holovko2017isotropic} the SPT2 approach was applied for the description of a bulk HS/HSC mixture.
We found from a comparison of the results obtained from the theory with computer simulations that the accuracy of the description 
of isotropic-nematic transition in a HS/HSC mixture becomes poor if the length of spherocylinders $L_2$ is not large enough. 
Since due to decreasing the length $L_2$ the region of isotropic-nematic phase transition shifts towards higher densities
it is reasonable to assume that the problem of the SPT2 approach with relatively small values of $L_2$ is related to the accuracy of the SPT theory 
for higher densities.

As it is known \cite{yukhnovski1980statistical} in the SPT for a HS fluid the thermodynamic properties coincide with the results obtained from the analytical solution of the Percus-Yevick integral equation, which is far from perfect at high fluid densities.
Therefore, it is necessary to introduce an efficient correction in our formalism. 
Such a simple correction is the Carnahan-Starling improvement \cite{kayser1978bifurcation}.

We propose the Carnahan-Starling correction for the case of a fluid in a matrix and present the equation of state
in the following way:
\begin{eqnarray}
\label{pressureCS1}
\frac{\beta P^{SPT2b-CS}}{\rho}=\frac{\beta P^{SPT2b}}{\rho}+\frac{\beta \Delta P^{CS}}{\rho},
\end{eqnarray}
where $\frac{\beta P^{SPT2b}}{\rho}$ is given by Eq.(\ref{pressureSPT2b}), $\frac{\beta \Delta P^{CS}}{\rho}$ is the Carnahan-Starling correction which can be defined as \cite{yukhnovski1980statistical}
\begin{eqnarray}
\label{pressCS1}
\frac{\beta \Delta P^{CS}}{\rho}=-\frac{\left(\eta/\phi_0\right)^3}{\left(1-\eta/\phi_0\right)^3}.
\end{eqnarray}

Carrying out a similar procedure described in the previous section we obtain the following expression for the free energy
\begin{eqnarray}
\label{energyCSandPL}
\frac{\beta F^{SPT2b-CS}}{V}=\frac{\beta F^{SPT2b}}{V}+\frac{\beta \Delta F^{CS}}{V},
\end{eqnarray}
where the first term $\frac{\beta F^{SPT2b}}{V}$ is given by Eq.(\ref{F2b}) and the second one is the following:
\begin{eqnarray}
\label{energyCS}
\frac{\beta \Delta F^{CS}}{V}=\rho\left[\ln(1-\eta/\phi_0)+\frac{\eta/\phi_0}{1-\eta/\phi_0}-\frac{1}{2}\frac{\left(\eta/\phi_0\right)^2}{\left(1-\eta/\phi_0\right)^2}\right]
\end{eqnarray}
Consequently, for the chemical potentials we have
\begin{eqnarray}
\label{muCSandPL}
\beta\mu_\alpha^{SPT2b-CS}=\beta\mu_\alpha^{SPT2b}+\beta\Delta\mu_\alpha^{CS},
\end{eqnarray}
where the first term $\beta\mu_\alpha^{SPT2b}$ is given by Eq.(\ref{chemSPT2b}) and the second one is
\begin{eqnarray}
\label{muCS}
\beta\Delta\mu_\alpha^{CS}=\ln(1-\eta/\phi_0)+\frac{\eta/\phi_0}{1-\eta/\phi_0}-\frac{1}{2}\frac{\left(\eta/\phi_0\right)^2}{\left(1-\eta/\phi_0\right)^2}
-\frac{\eta_\alpha/\phi_0}{x_\alpha}\frac{\left(\eta/\phi_0\right)^2}{\left(1-\eta/\phi_0\right)^3}.
\end{eqnarray}

Another point for improvement of the SPT2b theory is related to the expression (\ref{Cp}) for the parameter $C$ in the nonlinear equation (\ref{nonlineareq})
for the singlet distribution function $f(\Omega)$. The expression (\ref{Cp}) for the parameter $C$ has two terms. First of them appearing due to 
the coefficient $a_2(\tau(f))$ and the second one comes from the coefficient $b_2(\tau(f))$ in the expression (\ref{F2b}) for the free energy. 
We propose the analogues for these two terms in the corresponding integral equation for the singlet distribution function from the consideration of the HSC fluid using the Parsons-Lee (PL) approach \cite{parsons1979nematic,lee1987numerical}.
It was shown that the first term, which appears from the coefficient $a_2(\tau(f))$ in the SPT2b approximation and in the PL approach are the same. But there is some difference in the second term.
It is easy to show that when we introduce parameter $\delta$ as a factor before the term with $\tau(f)$ in the coefficient $b_2(\tau(f))$ it is possible to have practically the same results for the description of isotropic-nematic transition from both the SPT2 and PL approaches. After the generalization of this result for the case of HS/HSC mixture in a HS matrix, we can rewrite the expression for $b_2(\tau(f))$ in the following form:
\begin{eqnarray}
\label{b2pPL1}
&&b_2(\tau(f))=\left[\left(\frac{3}{4}s_1+\frac{3}{2}k_1\right)\frac{\eta_1}{\eta}
+\left(\frac{3(2\gamma_2-1)}{3\gamma_2-1}
+\delta\frac{3(\gamma_2-1)^2\tau(f)}{3\gamma_2-1}\right)\frac{\eta_2}{\eta}
-\frac{p'_{0\alpha}}{\phi_0}-\frac{1}{2}\frac{p'_{0\lambda}}{\phi_0}\right] \nonumber \\
&&\times \left(3k_1\frac{\eta_1}{\eta}+\frac{6\gamma_2}{3\gamma_2-1}\frac{\eta_2}{\eta}-\frac{p'_{0\lambda}}{\phi_0}\right).
\end{eqnarray}
Using the Parsons-Lee theory within the framework of Onsager's investigation for sufficiently long HSC particles we derive $\delta=\frac{3}{8}$ \cite{lee1987numerical}. As a result, we obtain a new expression for $C$:
\begin{eqnarray}
\label{Cp-PL}
&&C=\frac{\eta_2/\phi_0}{1-\eta/\phi_0}
\bigg[\frac{3(\gamma_2-1)^2}{3\gamma_2-1}\left(1-\frac{1}{2}\frac{p'_{0\lambda}}{\phi_0}\right) \nonumber \\
&&+\frac{1/\phi_0}{1-\eta/\phi_0}\frac{(\gamma_2-1)^2}{3\gamma_2-1}\delta
\left(3k_1\eta_1+\frac{6\gamma_2}{3\gamma_2-1}\eta_2-\frac{p'_{0\lambda}}{\phi_0}\eta\right)\bigg].
\end{eqnarray}


\section{Results and discussions} \label{sec:results}
\begin{figure}[htb]
	\begin{center}
		\includegraphics[width=0.55\textwidth]{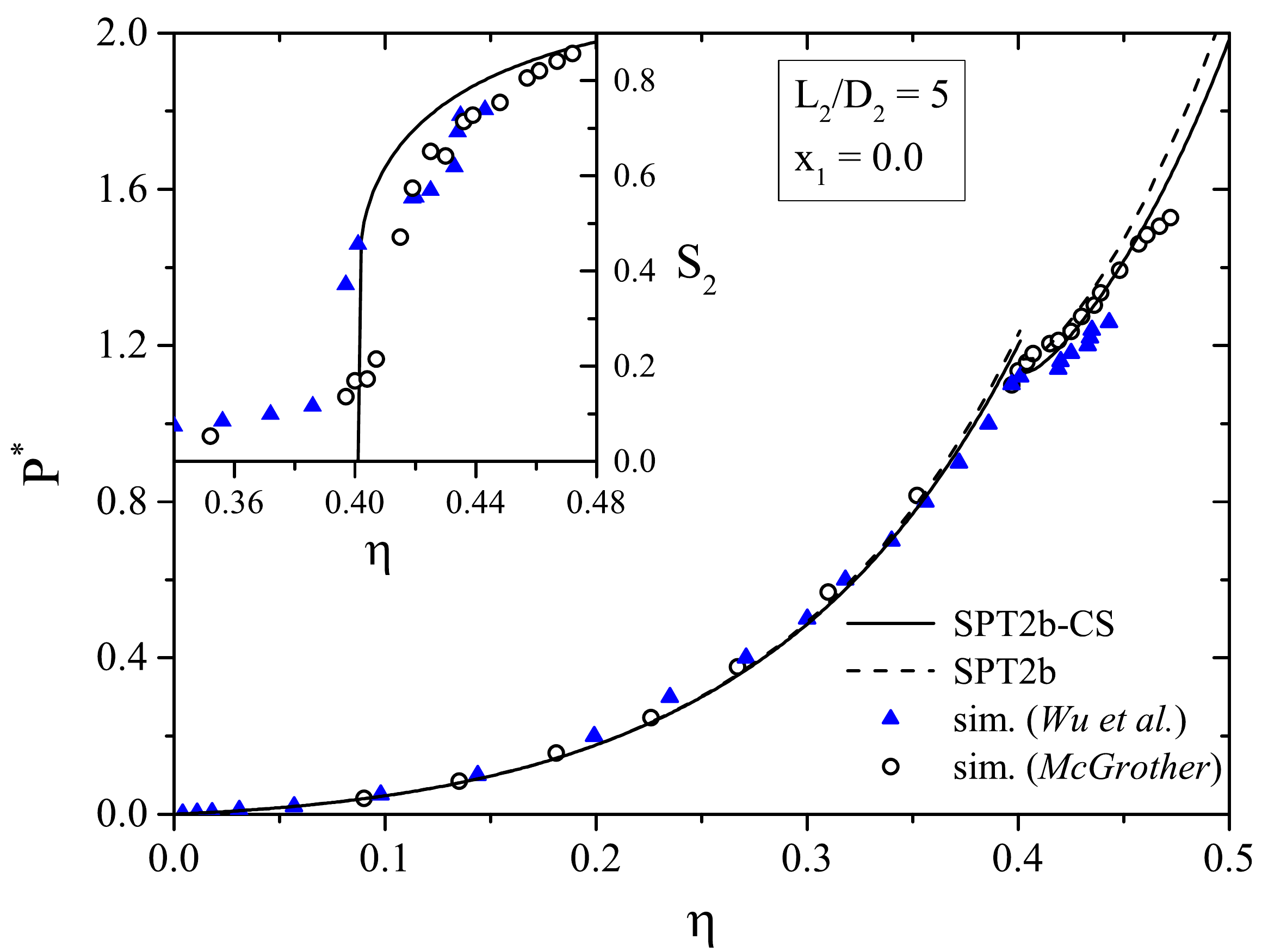}\\
		\includegraphics[width=0.55\textwidth]{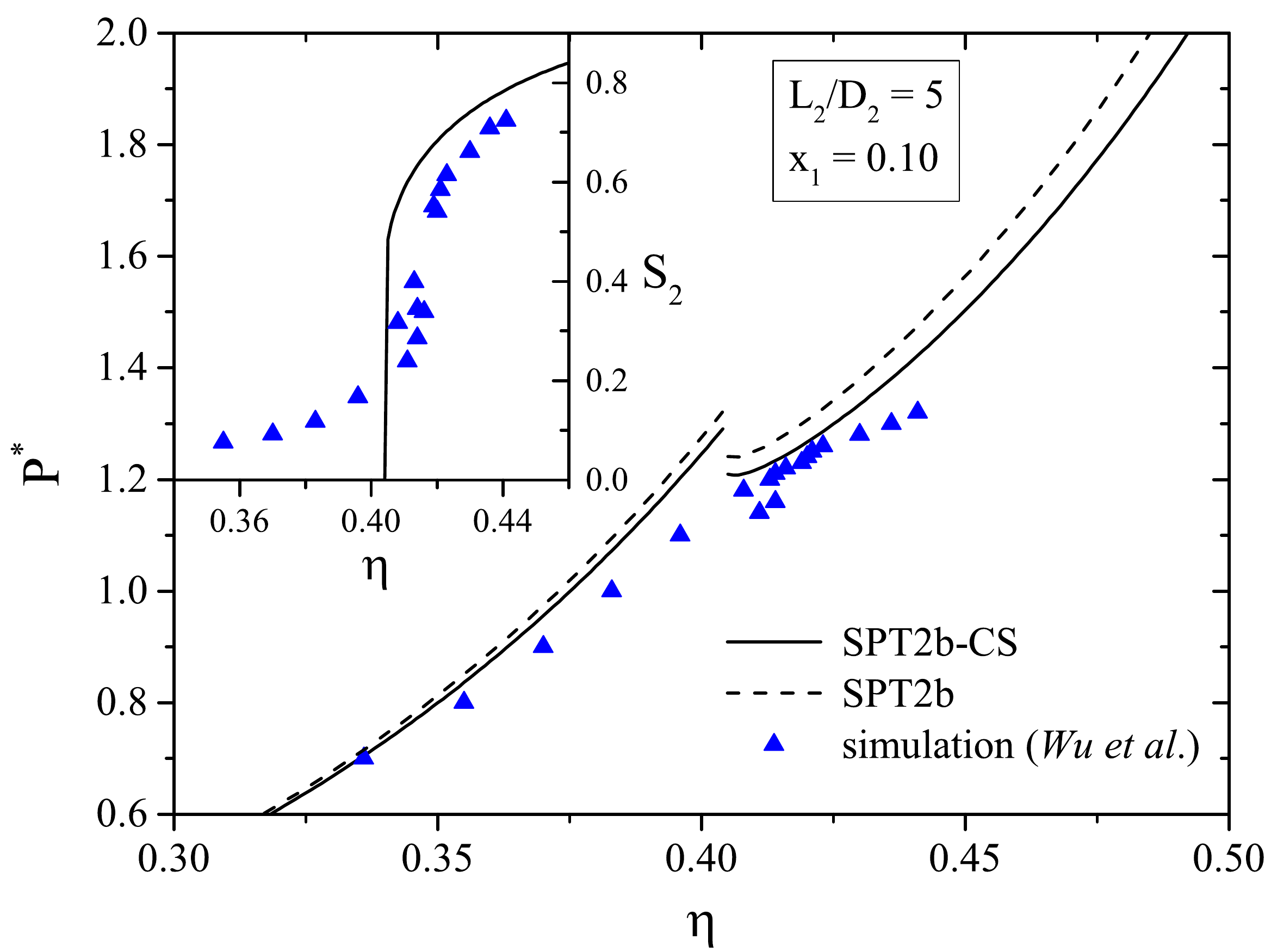}\\
		\includegraphics[width=0.55\textwidth]{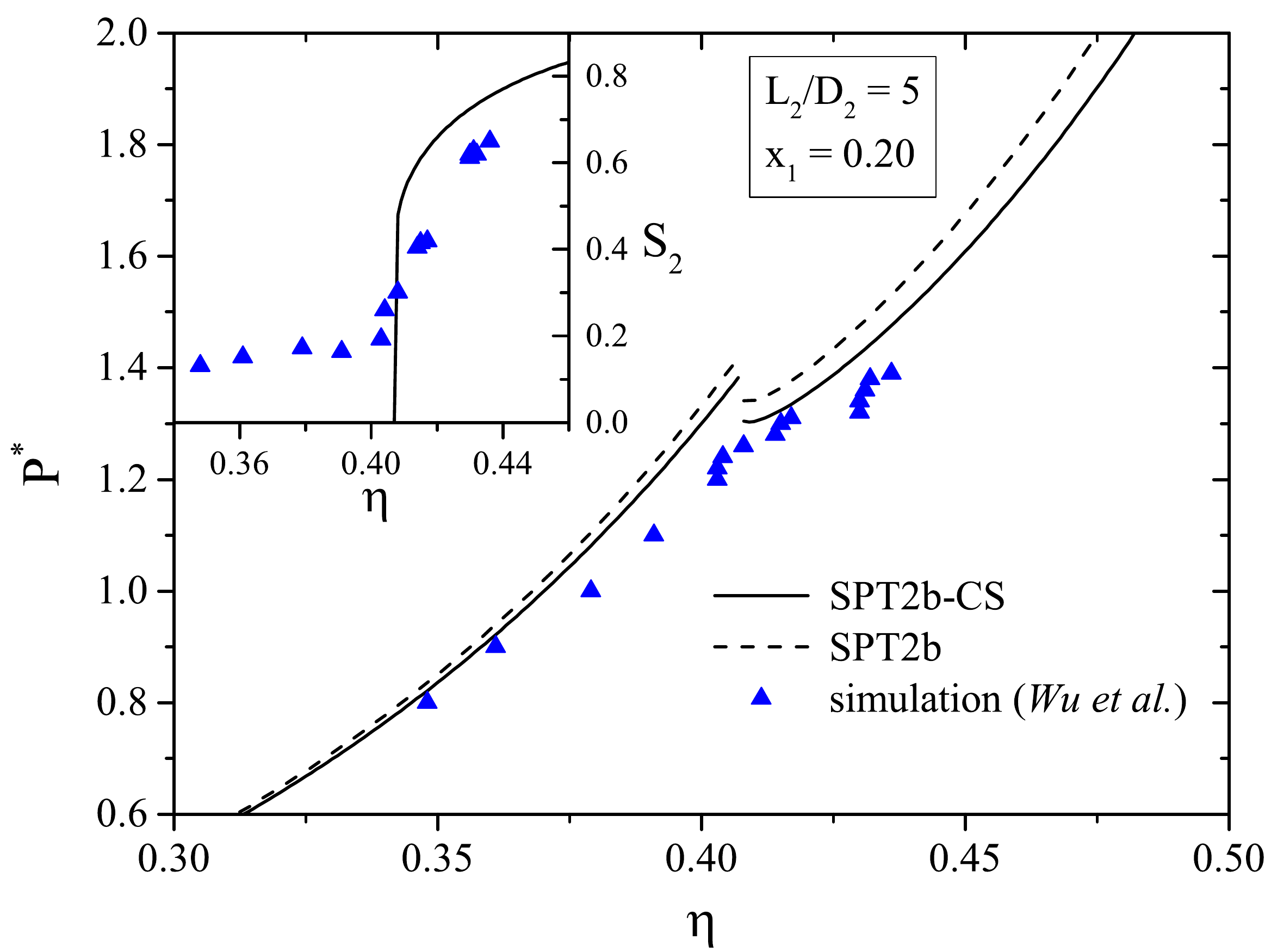}
		\caption{\label{fig:PetaL5} The equation of state for HS/HSC mixtures
			with $L_{2}/D_{2}=5$ and $D_{1}=D_{2}$ at different molar fractions of HS, $x_{1}$. Lines represent results obtained within the SPT2b-CS (solid) and SPT2b
			approaches (dashed). The insets represent
			the order parameter of HSC, $S_{2}$, depending on the packing fraction
			of HS/HSC mixture, $\eta$. Simulation results taken from the literature are presented as symbols: circles\cite{mcgrother1996} 
			and triangles\cite{wu2015orientational}.}
	\end{center}
\end{figure}

The approach presented in the previous section allows us to describe thermodynamic properties of
binary HS/HSC mixtures confined in a disordered porous medium of a HS matrix. It is important to 
ascertain that this theory can provide correct results for the thermodynamic quantities of considered systems
and that it is able to predict accurately the position of I-N transition occurring in a HSC component.
Therefore, using the SPT2 approach we have calculated the pressure of HS/HSC mixture $P$ as a function of its packing fraction $\eta$.
In Figs.~\ref{fig:PetaL5} we compare the obtained results for bulk HS/HSC mixtures at different HS concentrations $x_{1}$ with the computer simulation data taken from the literature\cite{mcgrother1996, wu2015orientational}. We consider the approximation SPT2b and its modification SPT2b-CS. As it was expected 
the SPT2b-CS coincides much better with the simulations than the SPT2b. Especially, it is seen for the case of mixtures ($x_{1}=0.1$ and $x_{1}=0.2$).
While for the pure HSC fluid ($x_{1}=0$) the difference between these two approximations is negligible. 
In this case the theoretical results perfectly fit the simulation data from \cite{mcgrother1996}. However, the worse coincidence 
with the simulations taken from \cite{wu2015orientational} is noticed, where the values of pressure obtained from the simulations become lower 
than theoretical ones at the packing fractions $\eta$ close to the I-N transition. This tendency also remains for the HS/HSC mixtures (Figs.~\ref{fig:PetaL5}) for the concentrations $x_{1}=0.1$ and $x_{1}=0.2$. Since for all considered concentrations $x_{1}$ 
the SPT2b-CS fits the simulation results for the pressure better than those obtained in the SPT2b, hereafter we use the SPT2b-CS approximation 
as the best one.
It should be noted that since the SPT2b-CS is based on the conceptions similar to the Parsons-Lee\cite{lee1987numerical} (PL) and many-fluid Parsons\cite{malijevsky2008many} (MFP) approaches, the results obtained for a bulk mixture are close
to that presented in \cite{wu2015orientational}, where the PL theory and MFP approach was used and compared. 
Also the order parameter behavior looks very similar to that predicted in \cite{wu2015orientational}, and one can observe that the I-N phase transition
occurs at the correct packing fraction $\eta$.

We apply the theory presented above to study the isotropic-nematic phase transition in a binary HS/HSC mixture 
confined in a matrix formed by a disordered HS particles. For this purpose we consider the coexistence between 
isotropic and nematic phases of a HSC component depending on the concentration of HS particles $x_{1}$ and 
the packing fraction of matrix particles $\eta_0$.

A basic description of the isotropic-nematic phase behaviour of HSC can be obtained from the bifurcation analysis
of the integral equation (\ref{nonlineareq}) for the singlet distribution function $f(\Omega)$. 
This equation has the same form as the corresponding equation obtained by \textit{L.~Onsager} \cite{onsager1949} for a pure HSC fluid in the
limit of $L_2\rightarrow\infty$ and $R_2\rightarrow0$, when the reduced density of fluid 
$c_2=\frac{1}{2}\pi\rho_2 L_2^2 R_2$ is fixed.
Using the bifurcation analysis of the integral equation (\ref{nonlineareq}) it was found that this equation 
has two characteristic points $C_i$ and $C_n$ \cite{kayser1978bifurcation}, which define the range of stability of considered mixture. 
The first point $C_i$ corresponds to the highest possible density of HSC in the stable isotropic state. 
The second point $C_n$ corresponds to the lowest density of the stable nematic state. 
For the Onsager model the minimization of free energy with respect to the singlet distribution function $f(\Omega)$
leads to the coexisting equations, the numerical solution of which provides the densities of coexisting isotropic and nematic phases of HSC fluid \cite{herzfeld1984highly,lekkerkerker1984isotropic,chen1993nematic}:
\begin{equation}
\label{CiandCnPorosity}
c_i=3.289,\qquad c_n=4.192.
\end{equation}

\begin{figure}[htb]
	\begin{center}
		\includegraphics[width=0.49\textwidth]{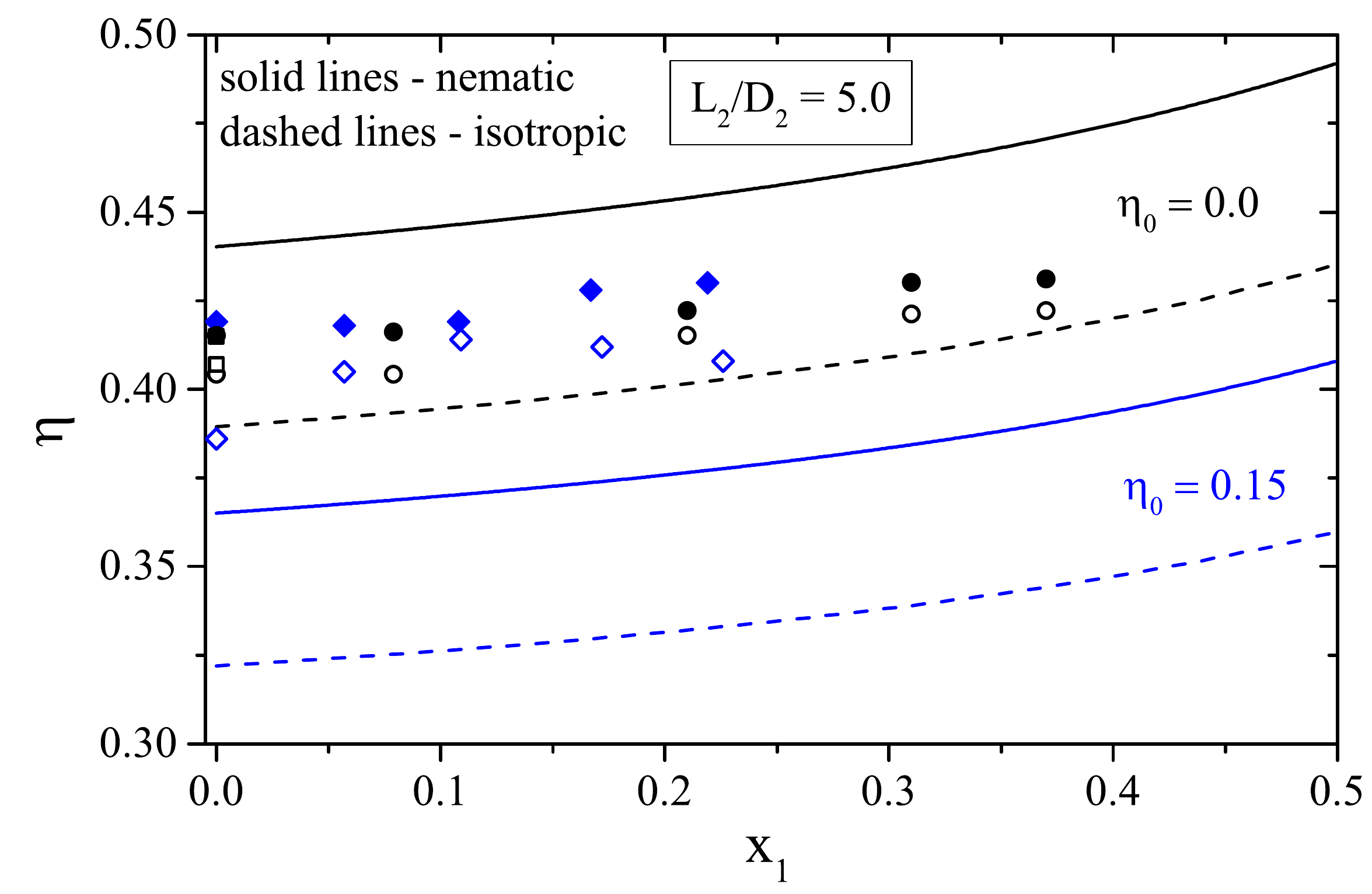}
		\includegraphics[width=0.49\textwidth]{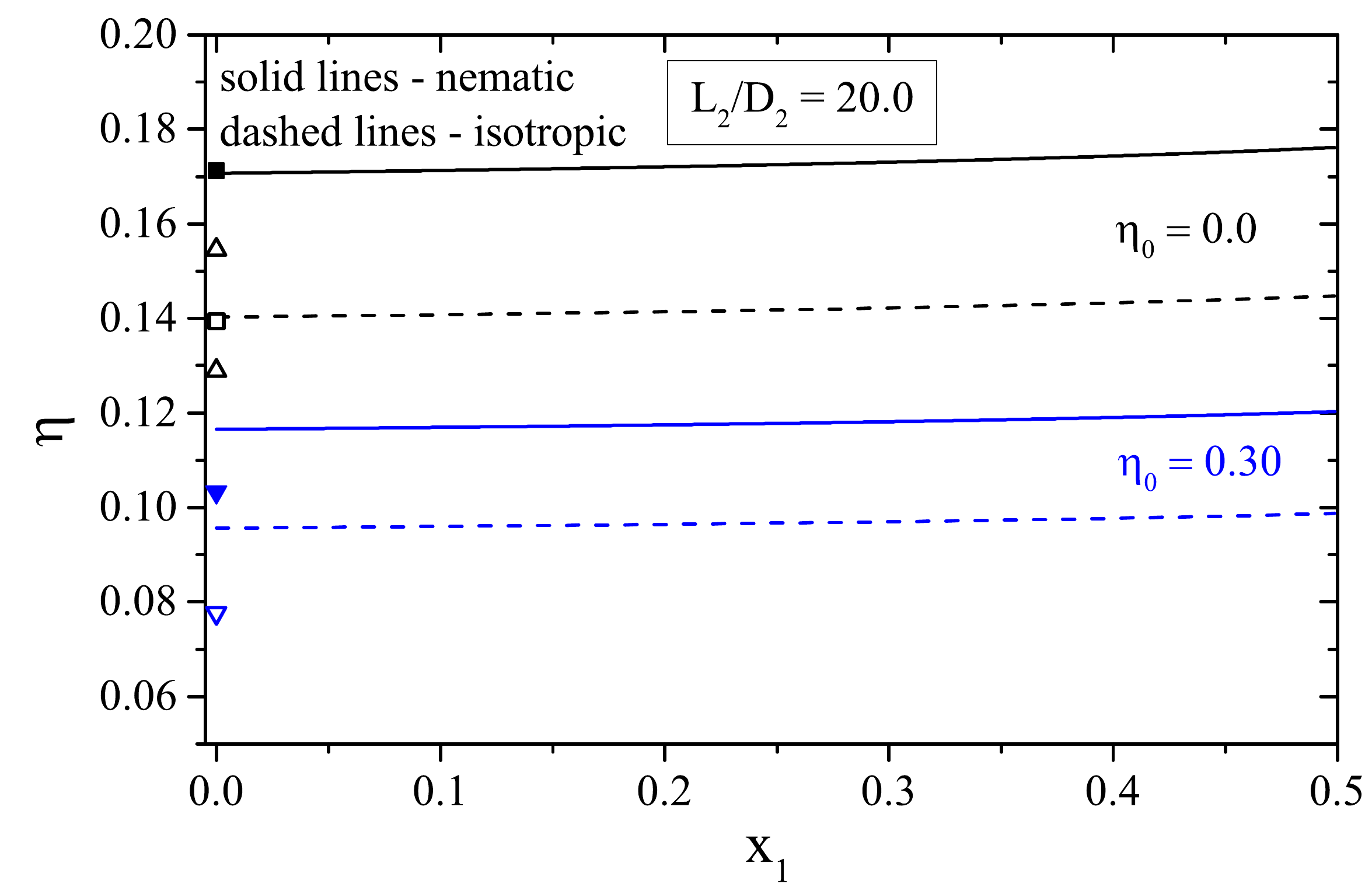}
		\caption{\label{fig:bifxL5} The results of bifurcation analysis. Isotropic-nematic coexistence curves of a HS/HSC mixture with $L_{2}/D_{2}=5$ (left panel) and $L_{2}/D_{2}=20$ in the bulk ($\eta_{0}=0.0$) and in a matrix
			of corresponding packing fraction $\eta_{0}$. 
			Sizes of spheres and matrix particles are equal to $D_{1}=D_{2}$ and $D_{0}=L_{2}$, respectively. 
			Simulation results taken from the literature are presented as symbols: circles\cite{lago2004crowding}, diamonds\cite{wu2015orientational}, squares\cite{frenkel1997} and triangles\cite{schmidt2004isotropic}. 
			Dashed lines and open symbols correspond to highest densities of isotropic phase. Solid lines and solid symbols
			denote lowest densities of nematic phase.}
	\end{center}
\end{figure}

For a HS/HSC mixture confined in the presence of porous medium we can use the Onsager limit and derive the expression for $C$ from (\ref{Cp-PL}) 
in the following form:
\begin{eqnarray}
\label{C_discussion}
C=\frac{c_2}{\phi_0} \frac{1}{1-\eta_1/\phi_0}.
\end{eqnarray}
It is immediately seen from the equation (\ref{C_discussion}) that the I-N transition 
of HSC particles confined in a matrix shifts towards lower densities of $c_{2}$ if the packing fraction of 
matrix $\eta_{0}=1-\phi_{0}$ increases. Similarly the HSC density $c_{2}$ is affected by increasing the
packing fraction of HS particles $\eta_{1}$.
In the general case when the parameters $L_{2}$ and $R_{2}$ are finite we can set
\begin{equation}
\label{CiandCnPorosityBig}
C_i=3.289,\qquad C_n=4.192,
\end{equation}
where $C_i$ and $C_n$ are determined from Eq.~(\ref{Cp-PL}).
This approach allows us to estimate the packing fractions of HS/HSC mixture at which
the isotropic and nematic phases of HSC component coexist.
Using the conditions (\ref{CiandCnPorosityBig}) we solve Eq.~(\ref{Cp-PL}) with respect to $\eta$,
where $\eta_{2}$ can be defined through $x_{1}$ as
\begin{equation}
\label{x2eta2}
\eta_{2}=\frac{\eta (1-x_{1}) V_{1}}{x_{1} V_{1} + (1-x_{1}) V_{2}},
\end{equation}
and $x_{1}$ is the concentration of HS particles in a HS/HSC mixture and $\eta$ is the total packing
fraction of this mixture.
Hence, the packing fractions of isotropic phase $\eta=\eta_{i}$ and nematic phase $\eta=\eta_{n}$ are calculated for a HS/HSC mixture in the bulk ($\eta_{0}=0$) and in a matrix.
In Fig.~\ref{fig:bifxL5} we present results of the bifurcation analysis for the I-N coexistence curves of a HS/HSC mixture in the $x_{1}\textrm{--}\eta$ plane. We consider two sizes of HSC particles $L_{2}/D_{2}=5$ (Fig.~\ref{fig:bifxL5}, left panel) and $L_{2}/D_{2}=20$ (Fig.~\ref{fig:bifxL5}, right panel). The sizes of HS particles are the same in the both cases and are
equal to the diameter of HSC particles ($D_{1}=D_{2}$). On the other hand the size of matrix particles is
set equal to the length of HSC particles ($D_{0}=L_{2}$). The values of the parameters for the considered model are taken according to that used in the simulation studies of \cite{lago2004crowding,wu2015orientational,schmidt2004isotropic}. Therefore, we can compare our
theoretical predictions with the simulation data. It is seen for the case of $L_{2}/D_{2}=5$ that 
the bifurcation analysis describes correctly the region of I-N phase transition
for a bulk HS/HSC mixture. However, the difference between coexistence densities $\eta_{n}$ and $\eta_{i}$
are much larger than that obtained from the computer simulations. On the other hand, for the long HSC particles 
($L_{2}/D_{2}=20$) the $\eta_{n}$ and $\eta_{i}$ are very close to the simulation results obtained for the bulk
HSC fluid in \cite{frenkel1997}. This witnesses that the accuracy of the bifurcation analysis improves with
increasing the length of HSC particles. Simultaneously, one can observe a rather poor coincidence of our
results with the simulation data of \textit{Schmidt} published in \cite{schmidt2004isotropic} for a pure HSC fluid in the bulk
and in a matrix. Surprisingly, both for the bulk HSC fluid and for the confined HSC fluid the simulations of \textit{Schmidt} \cite{schmidt2004isotropic} provide
somewhat lower results than the theory does. However, it can be related to the statistical error of simulations.
From the qualitative comparison of the coexistence curves shown in Fig.~\ref{fig:bifxL5} for the
bulk and confined HS/HSC mixtures one can notice that the packing fractions of isotropic and nematic phases
decrease with increasing the packing fraction of matrix, while the increase of HS concentration leads
to increasing of $\eta$. This trend is observed from both the bifurcation analysis and simulations.
Also the coexistence region is slightly narrower in the case of confined mixture than in the bulk case.

Although the bifurcation analysis gives us a rather good description of the isotropic-nematic transition
in a certain range of parameters, it remains rather limited, since it is not able to provide the whole phase
diagram for a HS/HSC mixture correctly even on a qualitative level. It was shown in \cite{cuetos2007useof} that at some
concentrations of HS particles the demixing processes occur in the coexisting phases leading to
the nematic phase rich in HSC particles and the isotropic phase rich in HS particles.
It follows from the conditions of thermodynamic equilibrium, according to which the pressure of HS/HSC mixture $P$ as well as the chemical potentials $\mu_{1}$ and $\mu_{2}$ of the both HS and HSC components in the isotropic phase should be the same as in the nematic phase:
\begin{equation}
\label{ThermodCoexistencePor}
P(\eta_{i},x_{1,i})=P(\eta_{n},x_{1,n}),\qquad \mu_{1}(\eta_{i},x_{1,i})=\mu_{1}(\eta_{n},x_{1,n}),
\qquad \mu_{2}(\eta_{i},x_{1,i})=\mu_{2}(\eta_{n},x_{1,n}),
\end{equation}
where $x_{1,i}$ and $x_{1,n}$ are the concentrations of HS particles in the isotropic and nematic phases, respectively.
This set of equations can be solved with respect to $\eta_{i}$, $\eta_{n}$ and $x_{i}$ at the fixed $x_{n}$.
Alternatively, we can fix $x_{i}$ and use $x_{n}$ as a variable. In any case, the calculated packing fractions and concentrations give us the dependencies of $\eta$ on $x_{1}$ in each of the phases, thus the coexistence curves can
be plotted in the $x_{1}\textrm{--}\eta$ plane.

Using the expressions for the pressure $P$ and the chemical potentials $\mu_{1}$ and $\mu_{2}$ 
derived by us in this study within the SPT2b-CS approximation (\ref{pressureCS1}) and (\ref{muCSandPL}) we have solved the equations (\ref{ThermodCoexistencePor}) numerically in combination with (\ref{nonlineareq}), where the definition
for $C$ used according to the expression (\ref{Cp-PL}). For this purpose the Newton-Raphson algorithm have been
applied. The packing fractions of HS/HSC mixture and HS concentrations in the coexisting isotropic and nematic phases have been calculated with 
the computational error less than $10^{-9}$.

\begin{figure}[htb]
	\begin{center}
		\includegraphics[width=0.7\textwidth]{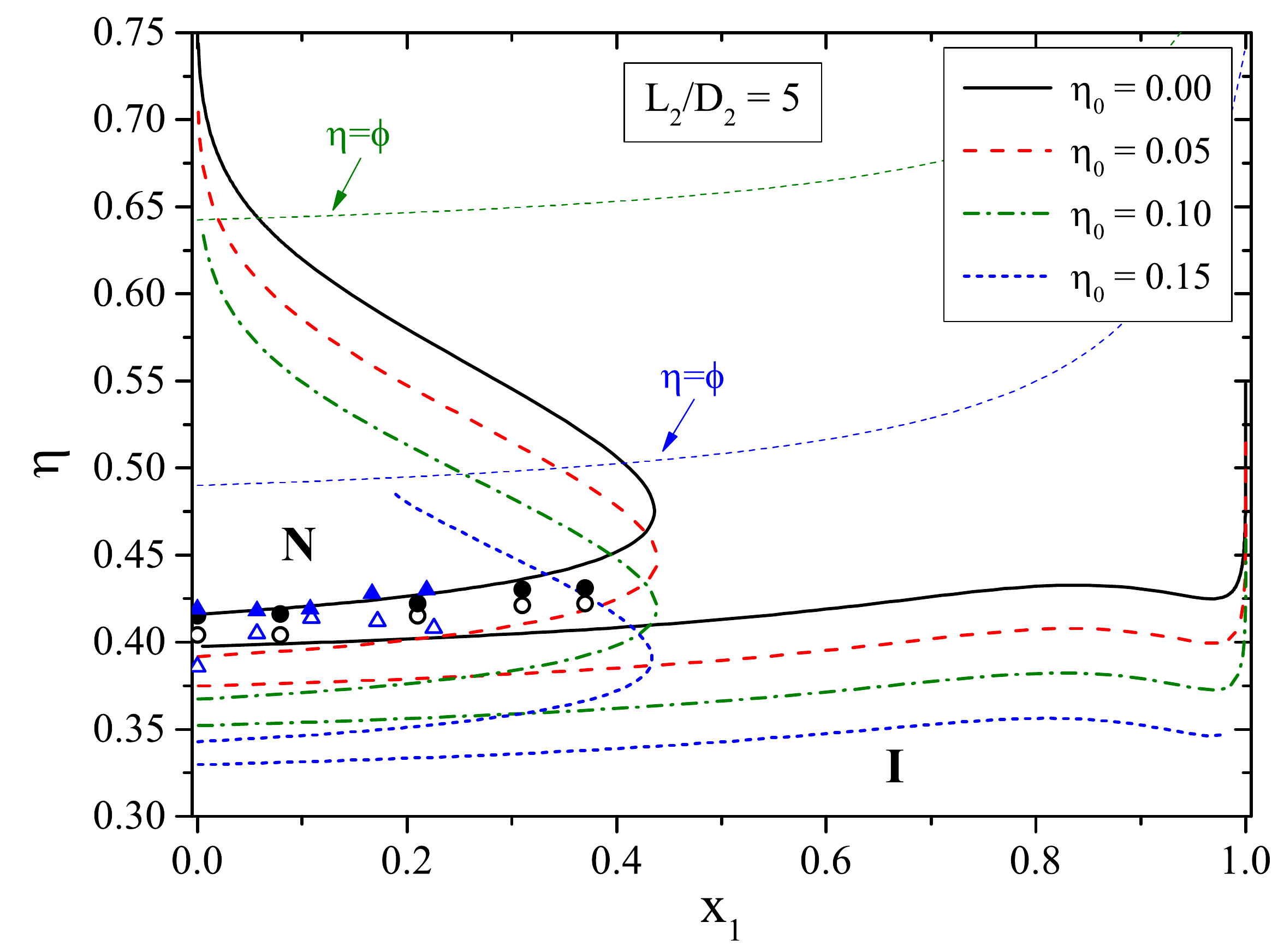}
		\caption{\label{fig:etaxcoexL5}
			Isotropic-nematic coexistence diagrams obtained within the SPT2b-CS approach for the HSC/HS mixture with $L_{2}/D_{2}=5$ and $D_{1}=D_{2}$ confined in matrices of different packing fractions $\eta_{0}=0.0-0.15$.
			Sizes of matrix particles equal to $D_{0}=L_{2}$.
			Lines represent results obtained within the SPT2b-CS approach.
			The simulation results (symbols) taken from \cite{lago2004crowding} and \cite{wu2015orientational} are shown as circles and triangles, respectively. 			
			Dashed lines and open symbols correspond to highest densities of isotropic phase. Solid lines and solid symbols
			denote lowest densities of nematic phase. Thin dashed lines pointed by arrows with $\eta=\phi$ labels indicate
			boundaries at which the mixtures reach their maxima of packing fraction $\eta$ in the corresponding matrices (green: $\eta_{0}=0.10$, blue: $\eta_{0}=0.15$).}
	\end{center}
\end{figure}

\begin{figure}[htb]
	\begin{center}
		\includegraphics[width=0.7\textwidth]{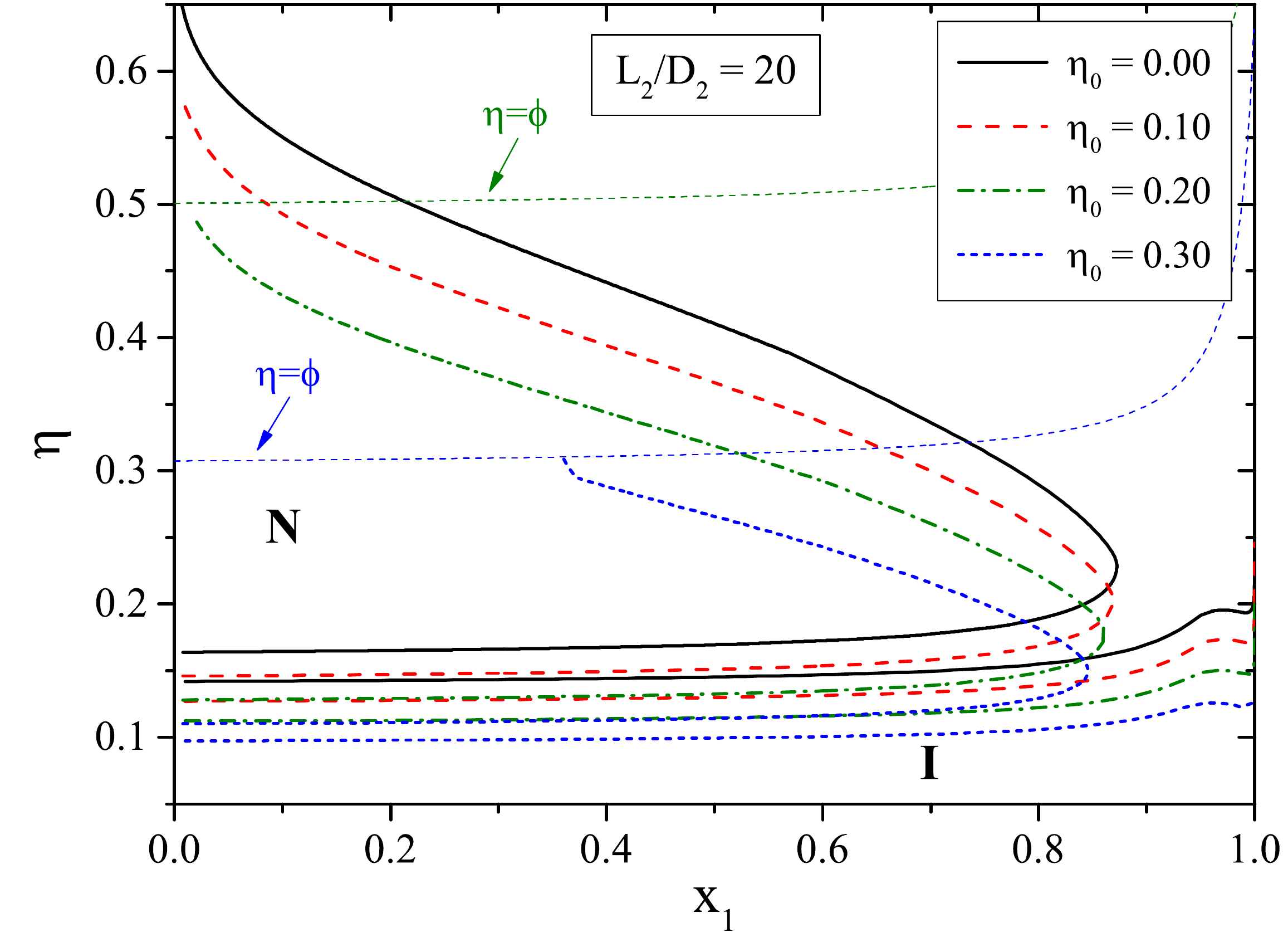}
		\caption{\label{fig:etaxcoexL20}
			Isotropic-nematic coexistence diagrams obtained within the SPT2b-CS approach for the HSC/HS mixture with $L_{2}/D_{2}=20$ and $D_{1}=D_{2}$ confined in matrices of different packing fractions $\eta_{0}=0.0-0.30$.
			Sizes of matrix particles equal to $D_{0}=L_{2}$.
			Lines represent results obtained within the SPT2b-CS approach.
			Dashed lines correspond to highest densities of isotropic phase and solid lines
			denote lowest densities of nematic phase. 
			Thin dashed lines pointed by arrows with $\eta=\phi$ labels indicate
			boundaries at which the mixtures reach their maxima of packing fraction $\eta$ in the corresponding matrices (green: $\eta_{0}=0.20$, blue: $\eta_{0}=0.30$).}
	\end{center}
\end{figure}

We present isotropic-nematic coexistence diagrams for confined HS/HSC mixtures with $L_{2}/D_{2}=5$ in Fig.~\ref{fig:etaxcoexL5} and with $L_{2}/D_{2}=20$ in Fig.~\ref{fig:etaxcoexL20}. Similarly as it was done above in the bifurcation analysis we set the size of HS particles equal to $D_{1}=D_{2}$ and the size of matrix particles is taken equal to $D_{0}=L_{2}$.
As it is seen the general shape of the coexistence curves of isotropic and nematic branches does not depend on the matrix porosity.
At low concentrations of HS the packing fraction $\eta$ increases monotonously with $x_{1}$ both in the isotropic and nematic phases. 
It should be noted that the concentration of $x_1$ in the nematic phase is always less than in the coexisting isotropic phase, 
and however this difference is not essential at lower packing fractions of nematic phase, it permanently increases with $\eta$.
Such a behaviour is observed till some certain point at which the system enters into the demixing regime, and the difference in HS concentrations
between two phases starts to increase much faster. This point corresponds to the concentrations $x_{1}\sim 0.44$ for $L_{2}/D_{2}=5$ 
and $x_{1}\sim 0.87$ for $L_{2}/D_{2}=20$.
It is also observed that at higher values of $x_{1}$ the nematic phase does not exist. 
Therefore, a further increase of $x_{1}$ in the isotropic phase
leads to decreasing of $x_{1}$ in the nematic phase. In this case the packing fraction of nematic phase increases permanently. 
On the other hand, in the isotropic phase the dependency of $\eta$ on $x_{1}$ is not monotonous, 
i.e. at high concentrations $\eta$ slightly decreases (at $x_{1}\sim 0.85$ for $L_{2}/D_{2}=5$ and $x_{1}\sim 0.97$ for $L_{2}/D_{2}=20$), and 
then it sharply rises when $x_{1}$ approaches unity. Simultaneously, in the nematic phase $x_{1}$ quickly decreases with
increasing $\eta$ and it tends to zero when the maximum packing fraction of HSC particles is reached.
This behaviour was described first in \cite{cuetos2007useof} for a bulk HS/HSC mixture by use of the Parsons-Lee approach. 
And now we can see that for confined HS/HSC mixtures the phase diagrams qualitatively repeat the bulk case,
although the I-N phase transition in the case of matrix presence occurs at lower packing fractions $\eta$. Nevertheless, we have also found some peculiarities in the phase behaviour of considered systems, which are 
related directly to confinement effects. Except shifting of the I-N coexistence region towards lower $\eta$, 
it have been noticed in Fig.~\ref{fig:etaxcoexL20} for $L_{2}/D_{2}=20$  that this region becomes narrower when the packing fraction of matrix increases. 
The same trend one can observe also for $L_{2}/D_{2}=5$ in Fig.~\ref{fig:etaxcoexL5}, however it is not well pronounced.
Another effect appears at the point from which the system enters into the demixing regime. For $L_{2}/D_{2}=5$ the demixing starts
at the same HS concentrations. On the contrary, for $L_{2}/D_{2}=20$ (Fig.~\ref{fig:etaxcoexL20}) this concentration varies depending on $\eta_{0}$ 
and it shifts towards lower values of $x_{1}$ with increasing $\eta_{0}$.

\begin{figure}[htb]
	\begin{center}
		\includegraphics[width=0.45\textwidth]{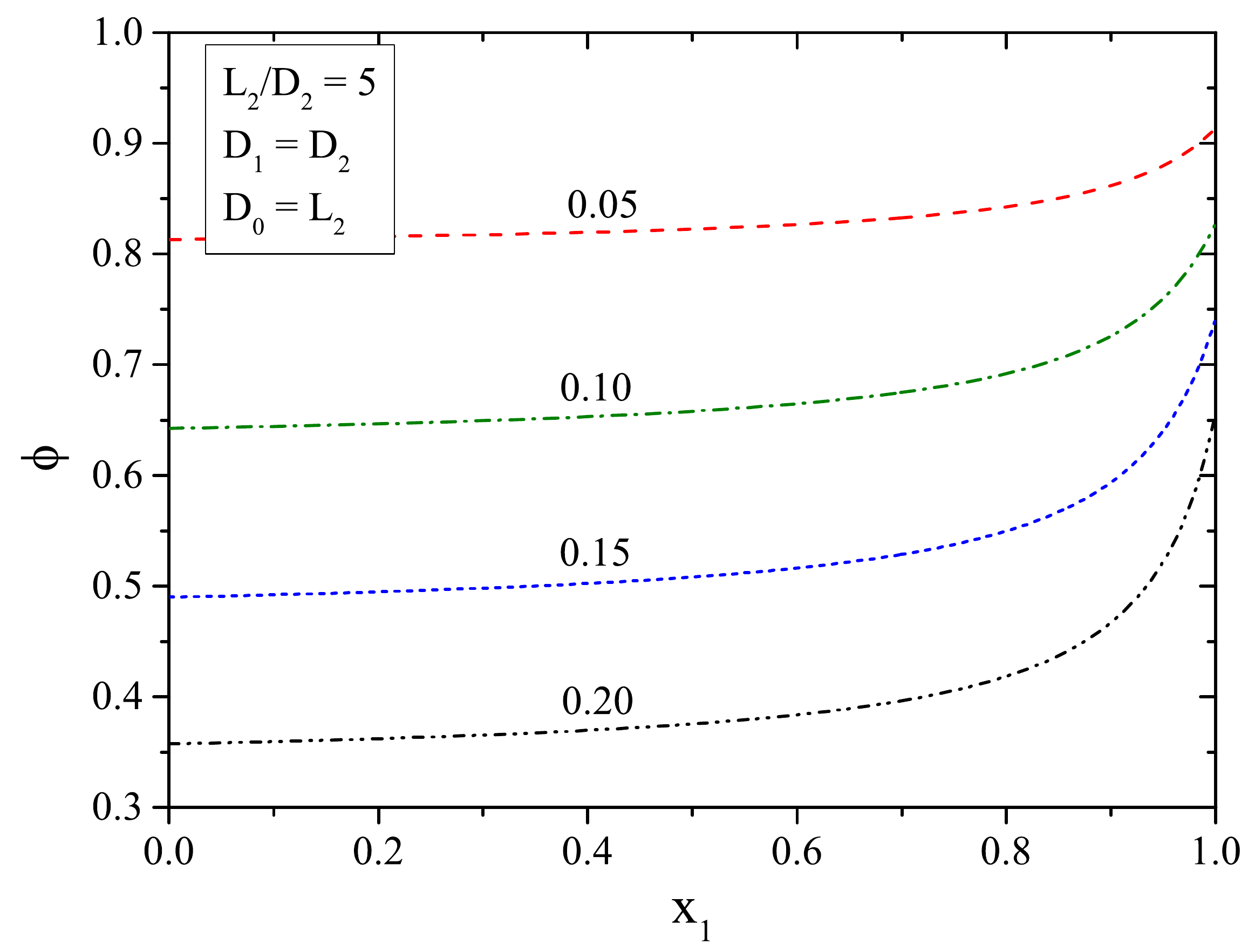}
		\includegraphics[width=0.45\textwidth]{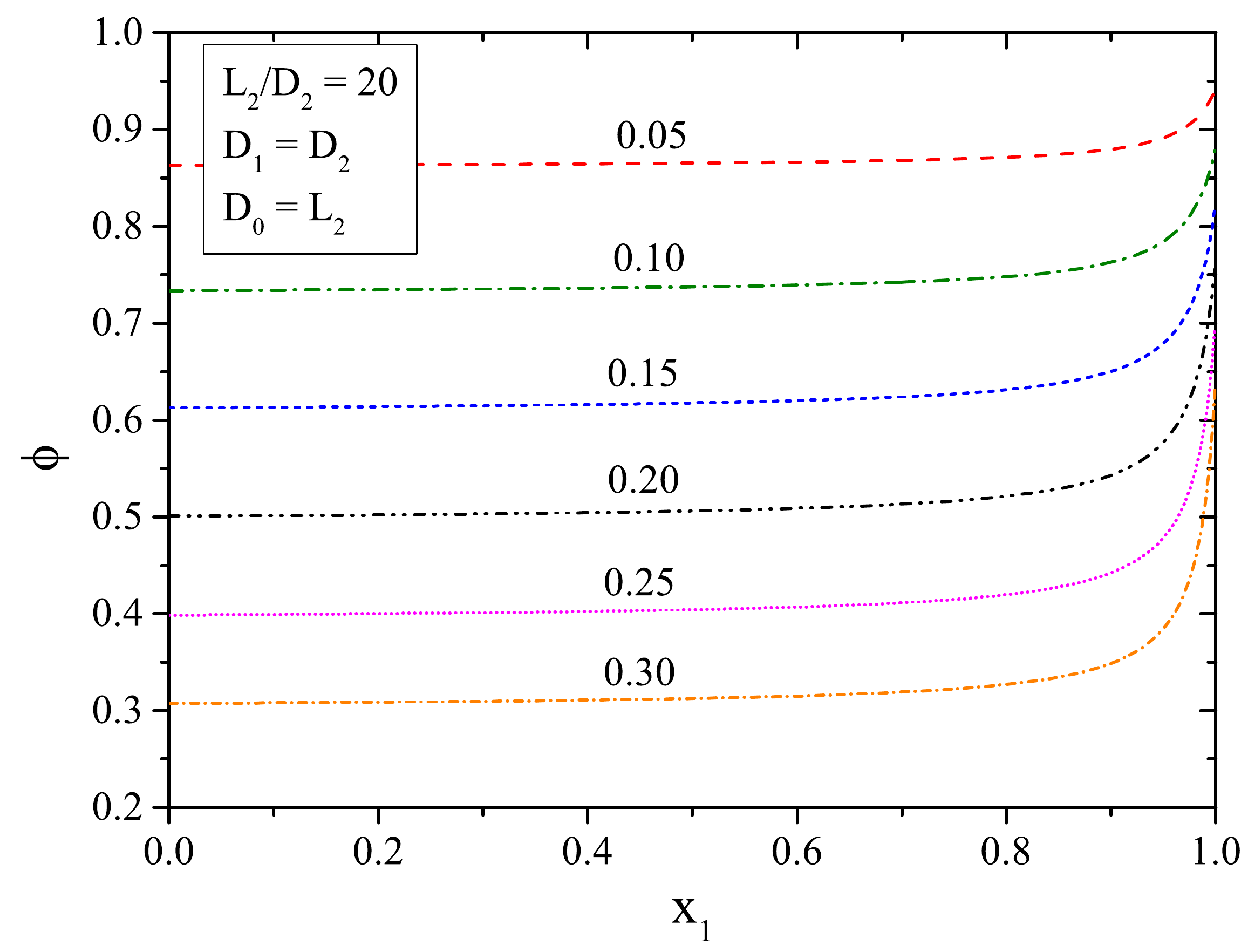}
		\caption{\label{fig:phixL5} The probe-particle porosity $\phi$ depending
			on the concentration of HS particles $x_{1}$ for the HS/HSC mixture with
			$L_{2}/D_{2}=5$ (left panel) and $L_{2}/D_{2}=20$ (right panel) confined in matrices of different packing
			fractions. Sizes of spheres and matrix particles are equal to $D_{1}=D_{2}$ and $D_{0}=L_{2}$, respectively.}
	\end{center}
\end{figure}

We should point out one more important effect, which is very specific to systems of particles confined in disordered matrices.
Especially, it concerns particles of strongly elongated shape, which cannot be packed efficiently in a disordered medium of matrix.
In our case such particles are spherocylinders and the maximum packing fraction of pure HSC particles is limited by the probe-particle 
porosity $\phi_2$ of the confining matrix (\ref{probeporosity02}). For pure spheres the probe-particle porosity 
is given by $\phi_{1}$ (\ref{probeporosity01}), which should be larger than $\phi_{2}$. 
For the HS/HSC mixture we have introduced the quantity $\phi$ (\ref{smallphip}), which is a combination 
of $\phi_{1}$ and $\phi_{2}$, and now we rewrite the expression (\ref{smallphip}) in terms of $\eta$ and $x_{1}$ as follows
\begin{equation}
\label{phi_on_eta_x1}
\phi=\left[x_{1}V_{1}+(1-x_{1})V_{2}\right]\left(\frac{x_{1}}{\phi_{1}}+\frac{(1-x_{1})}{\phi_{2}}\right)^{-1}.
\end{equation}
Using the obtained expression (\ref{phi_on_eta_x1}) and taking into account the condition for the packing fraction of HS/HSC mixture 
in a matrix, $\eta<\eta_{max}\equiv\phi$, we can predict a boundary line which cannot be crossed towards higher $\eta$ or lower $x_{1}$. 
In Figs.~\ref{fig:phixL5} we show some examples for the dependencies of $\phi$ on the concentration of HS particles $x_{1}$.
As it was expected the maximum packing fraction for HSC particles is less than for HS particles ($\phi(x_{1}=0)<\phi(x_{1}=1)$).
The dependencies are monotonous, thus for mixtures of HS/HSC the maximum packing fraction gets intermediate values.
These results explain a reason why some of coexistence curves stop at a certain point in Figs.~\ref{fig:etaxcoexL5} and \ref{fig:etaxcoexL20}
and cannot reach $x_{1}=0$ in the nematic phase at higher packing fractions.
For instance, this situation appears for HS/HSC mixtures in matrices of packing fractions $\eta_{0}=0.1$ and $0.2$ 
in the case of $L_{2}/D_{2}=5$, and for $\eta_{0}=0.2$ and $0.3$ in the case of $L_{2}/D_{2}=20$.

\begin{figure}[htb]
	\begin{center}
		\includegraphics[width=0.65\textwidth]{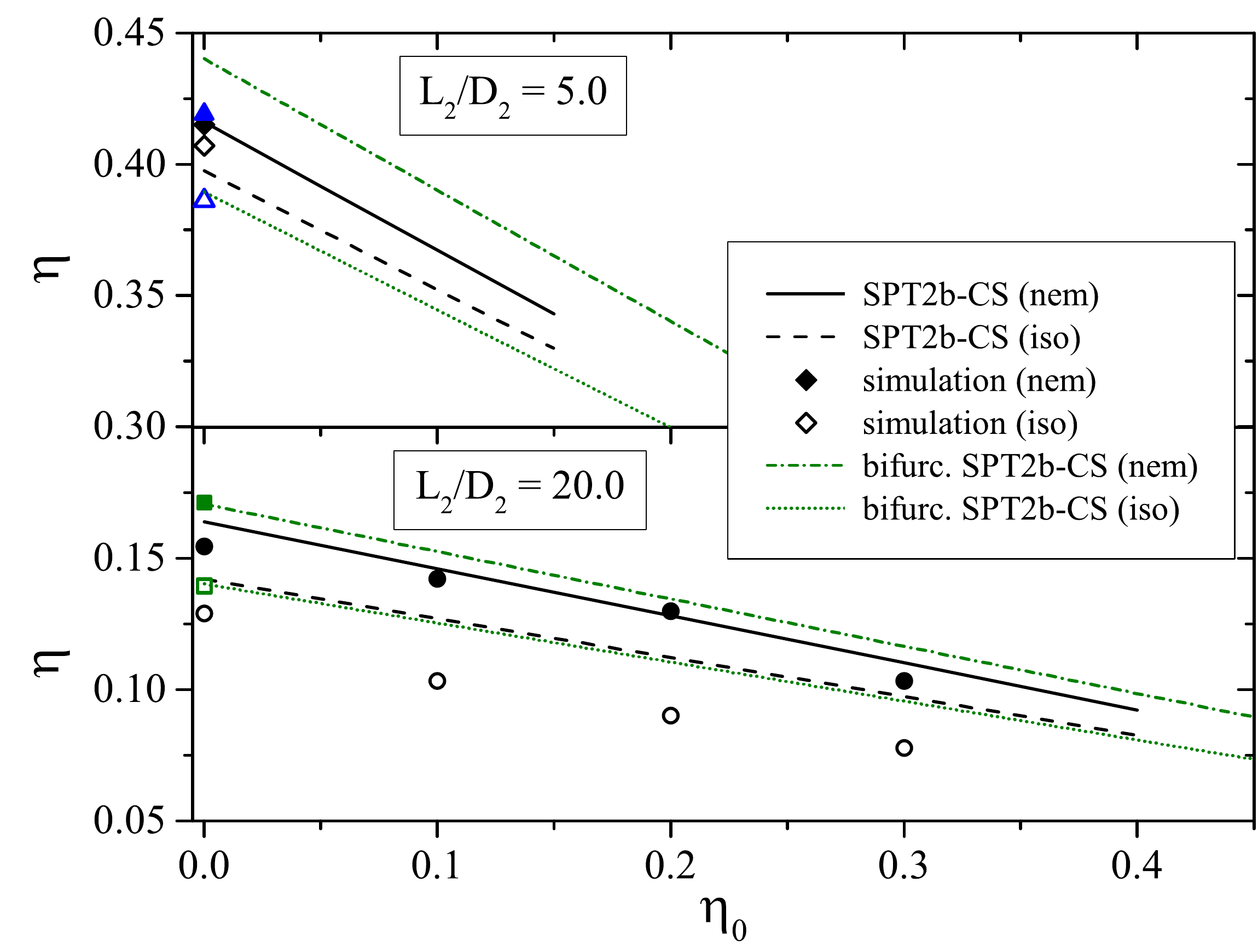}
		\caption{\label{fig:PxcoexL5} Isotropic-nematic coexistence curves of a pure HSC fluid with $L_{2}/D_{2}=5$ (upper panel) and $L_{2}/D_{2}=20$ (lower panel) confined in matrices of different packing fractions $\eta_{0}=0.0-0.30$.
			Solid and dashed lines denote results obtained within the SPT2b-CS approach.
			Dash-dotted and dotted lines correspond to the results of bifurcation analysis.
			Simulation results taken from the literature are presented as symbols: circles\cite{schmidt2004isotropic}, diamonds\cite{mcgrother1996}, squares\cite{frenkel1997} and triangles\cite{wu2015orientational}. }
	\end{center}
\end{figure}

Finally, we consider the I-N coexistence curves for the same system parameters 
as we used above ($L_{2}/D_{2}=5$ and $L_{2}/D_{2}=20$), but this time the results are shown in the $x_{1}\textrm{--}\eta_0$ plane.
We should note that to the best of our knowledge no simulation data for HS/HSC mixtures in disordered matrices are published for this moment.
However, one can find simulations for a pure HSC fluid confined in a matrix presented in \cite{schmidt2004isotropic}. 
Therefore, in Fig.~\ref{fig:PxcoexL5} we make an additional test of our theory by comparing our results with simulations in the case of $x_{1}=0$.
As it is seen for $L_{2}/D_{2}=5$ the approach proposed in our study describes well the nematic branch, while it leads to some overestimation for the isotropic curve. Similar tendency have also been observed for a bulk HS/HSC mixture in Fig.~\ref{fig:etaxcoexL5}, where we compare the I-N coexistence curves with the simulation results of \cite{lago2004crowding,wu2015orientational}.

\section{Conclusions}
The scaled particle theory previously generalized for the description of thermodynamical properties of a hard sphere fluid in a disordered 
porous matrix is extended to the case of a binary mixture of hard spherical colloids and spherocylinder particles. 
The analytical expressions for the pressure of the mixture and the chemical potentials of hard spheres and 
hard spherocylinders are obtained. For the correct description of thermodynamic properties the Carnahan-Starling-like and Parsons-Lee-like
corrections are introduced. The nonlinear integral equation for the orientation singlet distribution
function is obtained from the minimization of the free energy of considered system.
From the bifurcation analysis of this equation the isotropic-nematic phase transition in this binary mixture
is investigated. 
Other investigation of phase transition is done on the basis of conditions of thermodynamic equilibrium. It is shown that the both approaches
correctly reproduce the general trends of isotropic-nematic transition at small concentrations of hard spheres. 
However, the thermodynamic approach predicts the demixing transition at high concentrations of hard spheres, what is not available in 
the bifurcation analysis. The effect of disordered matrix on the isotropic-nematic and demixing transitions are discussed in terms
of the total packing fraction of mixture and the concentration of hard spheres for each of the coexisting phases.

It is shown that the increase of the packing fraction of matrix particles (decrease of porosity) shifts the region of isotropic-nematic
coexistence towards lower packing fraction of a mixture. Simultaneously, this region gets narrower. For the case of long spherocylinders
we have observed that the system enters the demixing regime at lower hard sphere concentrations if the packing fraction of matrix higher.
We found the existence of boundary $\eta--\phi$ at which the mixture reaches their maximum of packing fraction $\eta$ in corresponding matrices. 
For the binary mixture without matrix it was shown that the phase behaviour predicted from the theory is in a good agreement with existing computer simulations data.


\begin{acknowledgement}
	This project has received funding from the European Unions Horizon 2020 research and
	innovation programme under the Marie Sk{\l}odowska-Curie grant agreement No 734276, and from the State Fund For
	Fundamental Research (project N F73/26-2017).
	Authors are also grateful to V.~Shmotolokha for helpful discussions and valuable comments.
\end{acknowledgement}

\bibliography{HS_HSC_mat}

\newpage

\newpage
\begin{figure}[htb]
	\begin{center}
	\includegraphics[width=0.65\textwidth]{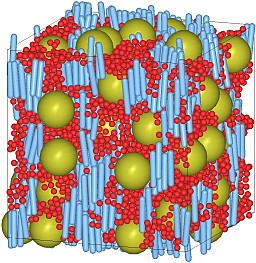}
	\caption{TOC Graphic}
	\end{center}
\end{figure}

\end{document}